\documentclass[pra,aps,showpacs,floatfix,twocolumn,superscriptaddress]{revtex4-1}
\usepackage{graphicx}
\usepackage[ansinew]{inputenc}
\usepackage{array}
\usepackage{color}
\usepackage{epsfig}
\usepackage{amsmath}
\usepackage{amsxtra}
\usepackage{amstext}
\usepackage{amssymb}
\usepackage{latexsym}
\usepackage{dsfont}
\usepackage{verbatim}

\newcommand{\bra}[1]{\left<#1\right|}
\newcommand{\ket}[1]{\left|#1\right>}

\newcommand{\eq}[1]{(\ref{eq:#1})}
\newcommand{\fig}[1]{Fig.~\ref{fig:#1}}

\begin{document}
\title{Hybrid optomechanical cooling by atomic $\Lambda$ systems}
\author{F. Bariani}
\thanks{These two authors contributed equally to the work.}
\affiliation{Department of Physics, College of Optical Sciences and B2 Institute, University of Arizona, Tucson, Arizona 85721, USA}
\author{S. Singh}
\thanks{These two authors contributed equally to the work.}
\affiliation{ITAMP, Harvard-Smithsonian Center for Astrophysics, Cambridge, Massachusetts 02138, USA}
\author{L.F. Buchmann}
\affiliation{Department of Physics, College of Optical Sciences and B2 Institute, University of Arizona, Tucson, Arizona 85721, USA}
\affiliation{Department of Physics, University of California, Berkeley, CA 94720, USA}
\author{M. Vengalattore}
\affiliation{Laboratory of Atomic and Solid State Physics, Cornell University, Ithaca, New York 14853, USA}
\author{P. Meystre}
\affiliation{Department of Physics, College of Optical Sciences and B2 Institute, University of Arizona, Tucson, Arizona 85721, USA}

\begin{abstract}
We investigate a hybrid quantum system consisting of a cavity optomechanical device optically coupled to an ultracold quantum gas. We show that the dispersive properties of the ultracold gas can be used to dramatically modify the optomechanical response of the mechanical resonator. We examine hybrid schemes wherein the mechanical resonator is coupled either to the motional or the spin degrees of freedom of the ultracold gas. In either case, we find an enhancement of more than two orders of magnitude in optomechanical cooling due to this hybrid interaction. Significantly, based on demonstrated parameters for the cavity optomechanical device, we identify regimes that enable the ground state cooling of the resonator from room temperature. In addition, the hybrid system considered here represents a powerful interface for the use of an ultracold quantum gas for state preparation, sensing and quantum manipulation of a mesoscopic mechanical resonator.  
\end{abstract}
\maketitle

\section{Introduction}
The control and manipulation of mesoscopic mechanical resonators by radiation pressure has fueled enormous interest in the use of optomechanical systems for applications to sensing, transduction and optomechanical information processing, as well as for foundational tests of quantum mechanics in the macroscopic domain \cite{OM_review}.  While cavity optomechanics has seen remarkable advances in recent years, the preparation of mesoscopic mechanical resonators in the quantum regime remains a significant challenge. In particular, the requirements for ground state cooling include the ability to isolate the resonator from environmental sources of dissipation as well as the realization of a cavity optomechanical system in the `resolved-sideband' regime \cite{OM_cooling}. Both these requirements become particularly challenging for optomechanical systems characterized by low frequency ($<$ 1 MHz) mechanical resonators.  Thus, for optomechanical systems operating in the optical domain, ground state cooling has, to date, only been demonstrated by the use of high frequency mechanical resonators and cryogenic cooling to reduce the thermal coupling to the environment \cite{Cleland,Teufel, Painter}. On the other hand, low frequency resonators are particularly compelling for a variety of sensing applications due to the long coherence times and large zero-point motion, and methods to cool such resonators to the quantum regime will enable a variety of technical applications as well as fundamental studies. 

A promising avenue to circumvent the limitations inherent to optomechanical cooling  is the use of an auxiliary quantum system that can enhance the effective optomechanical response of the resonator. Several proposals involving such hybrid schemes \cite{hybrid2014} and other approaches~\cite{PKU}  have been advanced to improve the performance of cavity cooling. Coupling a cold atomic ensemble to the mechanical resonator is of particular interest due to the precise control, wide tunability and strong optical interactions exhibited by atomic systems \cite{atom_cavity,treutlein-genes,bowen,genes}. 

\begin{figure}[b]
\begin{center}
\includegraphics[width=\columnwidth]{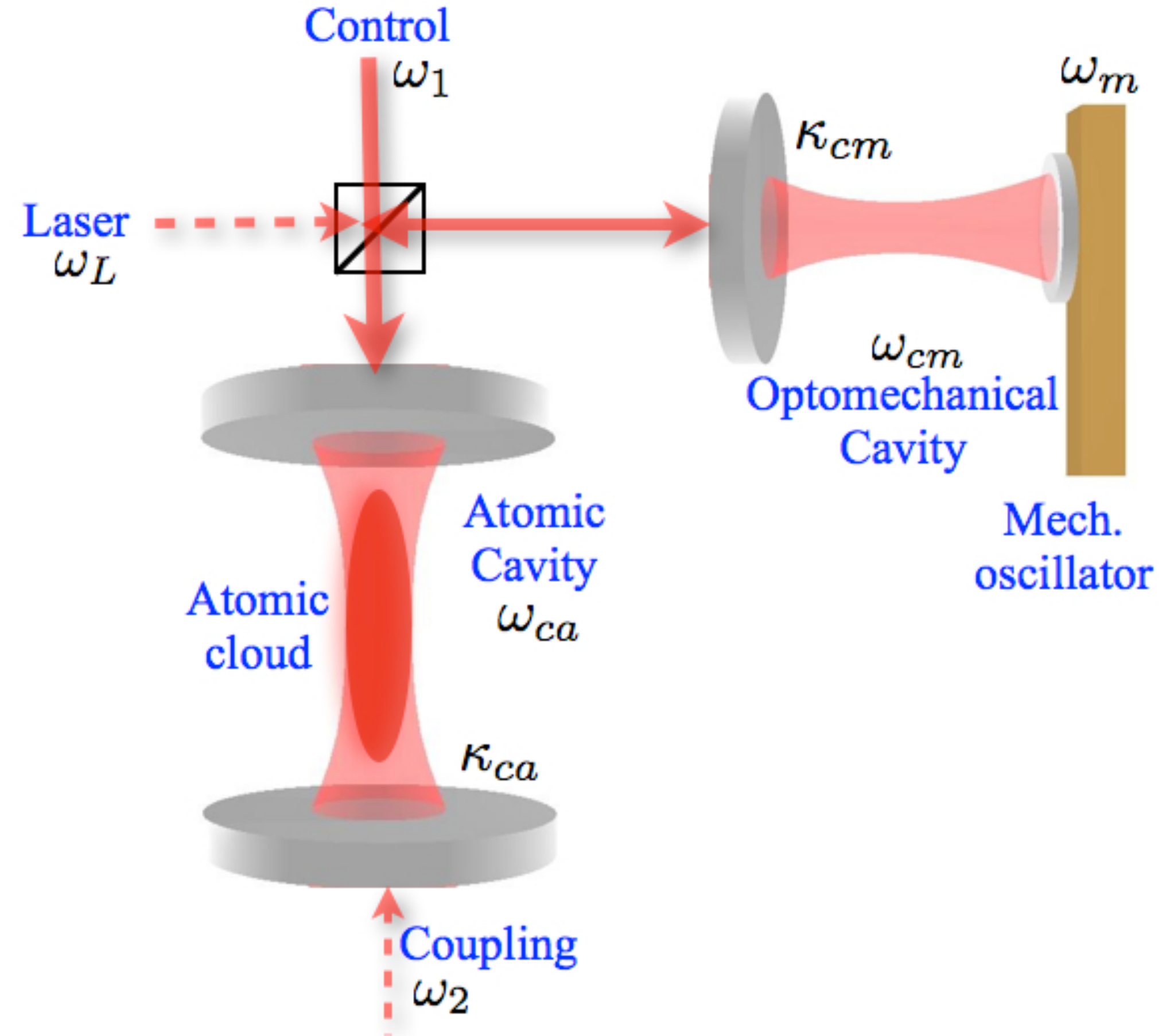}
\caption{(Colors online). Hybrid optomechanical setup with atoms. The cavities containing an atomic ensemble and an optomechanical element are coupled by an optical field that interacts with both the systems. }
\label{fig:AtomOM}
\end{center}
\end{figure}

This article investigates the hybrid optical interaction between a cavity optomechanical system and an ensemble of ultracold atoms. We consider the situation where the mechanical resonator and the ultracold gas are embedded within distinct optical cavities that interact via a common `coupling' laser (see Fig. 1). In addition to alleviating technical constraints, this modular approach also represents a powerful method to combine the coherence, sensitivity and tunability of the atoms with the robustness and scalability of the optomechanical device. 

As discussed below, the atomic gas is described as a three-level $\Lambda$ system whose optical response to the coupling light field can be modified by a strong `control' field. Such $\Lambda$ configurations in atomic ensembles exhibit narrow resonances due to the competition between dissipation and dispersion. We consider specifically two well-known and complementary approaches towards that goal, Electromagnetically Induced Transparency (EIT) \cite{EIT_review} and Recoil Induced Resonances (RIR)~\cite{Grynberg1992, Grynberg1994, Moore1998, Vengalattore2005, Hafezi2008}. In either case, and for experimentally demonstrated values of the quality factor and of the $Q \times f$ product for the mechanical resonator \cite{Chakram2014} and the cavity optomechanical system, we find a broad and robust range of parameters for which the mechanical resonator can be cooled to its quantum ground state from room temperature. 

In EIT, quantum interference between different excitation pathways to the same atomic level induces a narrow transparency window for the propagation of a weak resonant field that can be controlled via an external `control' laser. The effect may be tuned by separating two lower-lying states via e.g. Zeeman splitting by means of an additional magnetic field. Intracavity EIT has already been proposed for cooling and entanglement in an optomechanical setup \cite{Genes2009,Genes2011}. Here, we explore a new regime in separate cavities and we obtain the surprising result of a very effective blue-sideband cooling.

In contrast, RIR is generated when the exchange of energy between the control and the coupling lasers is resonant with the transition between different atomic momentum states. This nonlinear effect can be tuned by changing the frequencies of the lasers or the intensity of the control field. In particular, long-lived momentum coherences, narrow resonance features and large gain in the coupling beam amplitude have been demonstrated \cite{Hafezi2008}. We show that this spectrally narrow gain feature enhances the asymmetry between the optical sidebands at the mechanical frequency, which in turn leads to efficient cooling.

This paper is organized as follows. In Section II we present the complete Hamiltonian for the system, comprising the optical, mechanical and atomic parts, as well as the coupling between the different cavities. We expand on the atomic Hamiltonian in  Section III, reviewing the theory for the atomic response and discussing the atomic susceptibilities for EIT and RIR. Section IV is devoted to the analysis of the optomechanical dynamics, with emphasis on two possible configurations: cascade and feedback coupling between the cavities. Finally, we draw some conclusions and we discuss the perspectives of this work in Section V. 

\section{Model Hamiltonian}
We consider two coupled cavities, one containing an atomic ensemble and another including a mechanical resonator.  We will refer to them as the atomic cavity and the optomechanical cavity, respectively. The two cavities are optically coupled to obtain an effective interaction between the atomic ensemble and the mechanical element. A possible realization of the system is sketched in Fig. \ref{fig:AtomOM}. The full Hamiltonian describing this setup is 
\begin{equation}
H=H_{\rm ca}+H_{\rm atom}+H_{\rm cm}+H_{OM}+H_{AM}+H_{\rm loss},
\label{eq:totalH}
\end{equation}
where  $ H_{\rm ca}, H_{\rm cm} $ describe the optical fields in the two cavities, $H_{AM}$ represents the coupling between the two cavities, $H_{OM}$ contains the optomechanical interaction, $H_{\rm atom}$ accounts for the atomic dynamics that will be discussed in detail in the following Section, and $H_{\rm loss}$ denotes the various (atomic, optical and mechanical) loss mechanisms.

The cavities have the decay rates $\kappa_\mathrm{ci}={\rm FSR}_i/\mathcal{F}_\mathrm{ci}$, where ${\rm FSR}_i =2 \pi c/2L_\mathrm{ci}$ is the free spectral range, $L_\mathrm{ci}$ being the length of the cavity and $\mathcal{F}_\mathrm{ci}$ the cavity finesse. Here the subscripts $i \in \{a,m\}$ denote the atomic and optomechanical cavity, respectively. We assume throughout that $\mathcal{F}_\mathrm{ca}\ll\mathcal{F}_\mathrm{cm}$. The dynamics in the atomic cavity is described by $H_{\rm atom} + H_{\rm ca}$, with
\begin{equation}
H_{\rm ca}=\hbar\omega_\mathrm{ca} \hat{a}^\dagger \hat{a} +i\hbar(\eta_a \hat{a}^\dagger-\eta_a^*\hat{a}),
\label{eq:HoptA}
\end{equation}
where $\hat{a}$ denotes the annihilation operator for the cavity mode, which acts as the coupling field. The cavity is driven by $\eta_a=\sqrt{P_{\rm in,a} \kappa_\mathrm{l,ca}/\hbar\omega_\mathrm{ca}}$, with $P_{\rm in,a}$ the power of the input coupling beam and $\omega_\mathrm{ca}$ the cavity resonance frequency and $\kappa_\mathrm{l,ca}$ indicates the coupling through the left (input) mirror of the atomic cavity. Its value depends on the type of cavity we are considering \cite{kappa}.

Similarly the Hamiltonian for the optomechanical cavity can be written as $H_{\rm cm}+H_{OM}$ where
\begin{equation}
H_{\rm cm}=\hbar\omega_\mathrm{cm} \hat{c}^\dagger \hat{c} +i\hbar(\eta_c \hat{c}^\dagger-\eta_c^*\hat{c}),
\end{equation}
and $\hat{c}$ is the annihilation operator for the cavity mode with frequency $\omega_\mathrm{cm}$, $\eta_c=\sqrt{P_{\rm in,c} \kappa_\mathrm{l,cm}/\hbar\omega_\mathrm{cm}}$, $P_{\rm in,c}$ being the input power. The optomechanical interaction is
 \begin{equation}
 H_{OM}=\hbar\omega_m \hat{b}^\dagger \hat{b}+\hbar g_0 \hat{c}^\dagger \hat{c}(\hat{b}+\hat{b}^\dagger),
 \end{equation}
where $\hat{b}$  annihilates phonons of the relevant mechanical oscillator mode, with frequency $\omega_m$, and $g_0$ is the single photon optomechanical coupling. Finally, we describe the coupling between the cavities by the Hamiltonian
\begin{equation}
H_{AM} = \hbar J (\hat{a}^{\dagger}\hat{c} + \hat{c}^{\dagger}\hat{a}),
\end{equation}
where $J$ is a the phenomenological constant. The technical details leading to its specific value depend on the specifics of the experimental setup, in particular the mode matching of the coupling field between the two cavities. Neglecting all coupling losses and assuming perfect mode matching, we set $J=\sqrt{\kappa_\mathrm{l,ca}\kappa_\mathrm{l,cm}}$~\cite{GardinerZollerbook}. 

\section{Optical response of atomic $\Lambda$ schemes}
\begin{figure}[htbp]
\begin{center}
\includegraphics[width=\columnwidth]{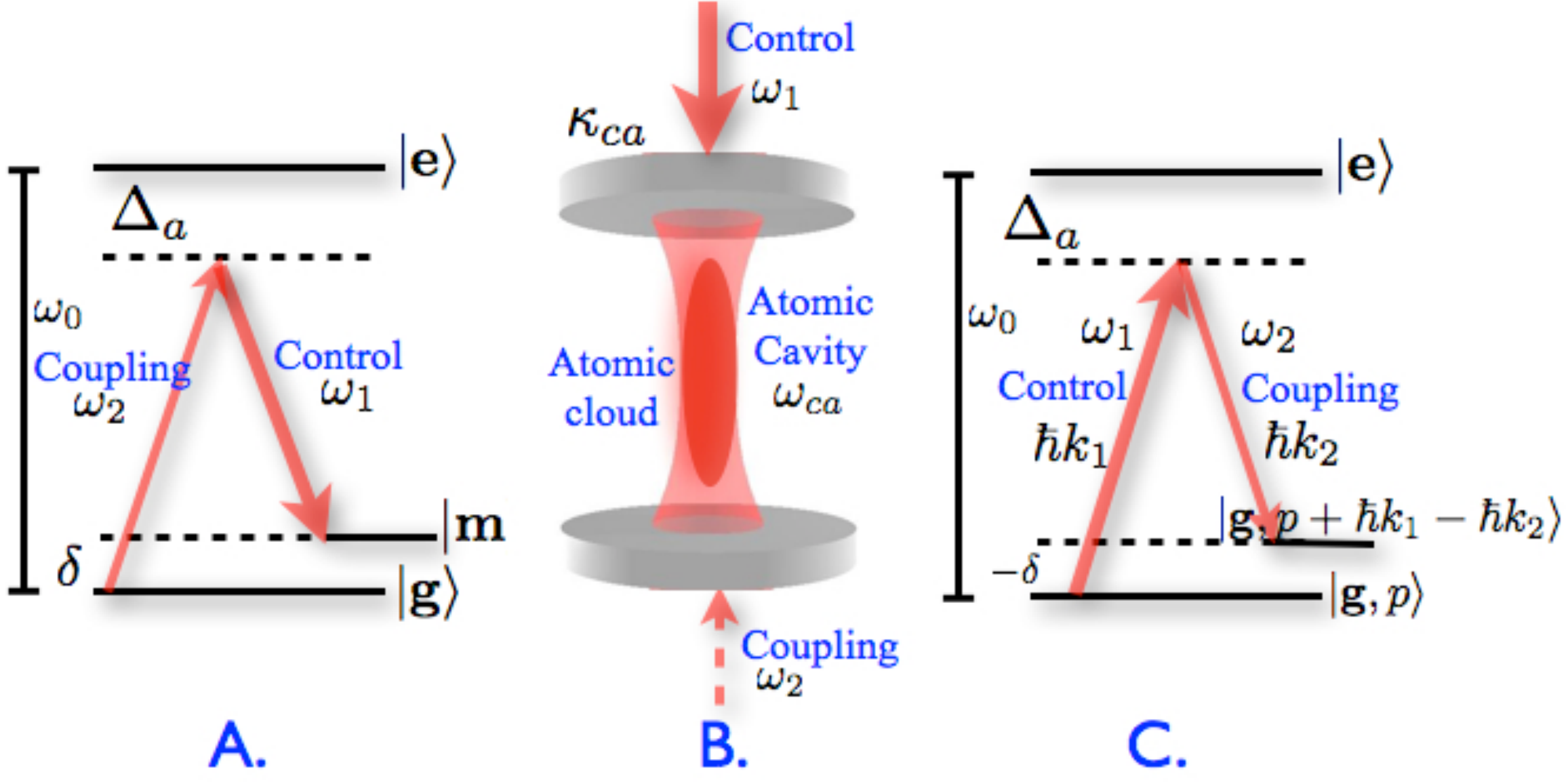}
\caption{(Colors online). Atomic level scheme for EIT (A) and RIR (C). Generic setup for an atomic ensemble in a single mode cavity (B). }
\label{fig:EITschemes}
\end{center}
\end{figure}

In this section we introduce two specific atomic $\Lambda$ configurations. The intent is to exploit their {\it narrow} and {\it tunable} spectral features to resolve the mechanical degree of freedom. The level diagrams are given in \fig{EITschemes}(A,C).  In both cases, we assume that the atomic ensemble is confined in an optical cavity, as shown in \fig{EITschemes}B. A strong (classical) control laser with frequency $\omega_1$ and a weak coupling laser with frequency $\omega_2$ interact with a chosen atomic transition of frequency $\omega_0$. To study the optical response of the atomic ensembles we consider an isolated atomic cavity with Hamiltonian
\begin{equation}
H=H_{\rm ca}+H_{\rm atom}+H'_{\rm loss},
\end{equation}
where $H'_{\rm loss}$ accounts for all relevant (atomic and optical) loss mechanisms. 

\subsection{Electromagnetically Induced Transparency}
In this scheme, sketched in \fig{EITschemes}A, we take advantage of the internal energy level structure of the atoms, in particular the splitting in Zeeman sublevels of a hyperfine manifold, where the energy difference between the states can be tuned via an external magnetic field $B_{\rm ext}$.  The two levels $|g\rangle$ and $|m\rangle$ are  Zeeman sublevels of the ground state manifold with energy difference $\Delta_m \mu_B g_F B_{\rm ext}$, where $\mu_B$ is the Bohr magneton, $g_F$ is the Land{\' e} factor and $\Delta_m$ is the difference in the magnetic quantum number. 
The state $|e\rangle$ is instead chosen from an excited manifold.

We assume a tightly trapped atomic sample, such that recoil effects are negligible. Furthermore, to ensure stable steady-state conditions we assume that the atomic ensemble is simultaneously cooled by Raman sideband cooling, so as to repopulate the ground state with high fidelity \cite{Patil2014} even in the presence of the coupling laser.

\begin{figure*}[ht]
\begin{center}
\includegraphics[width = \textwidth]{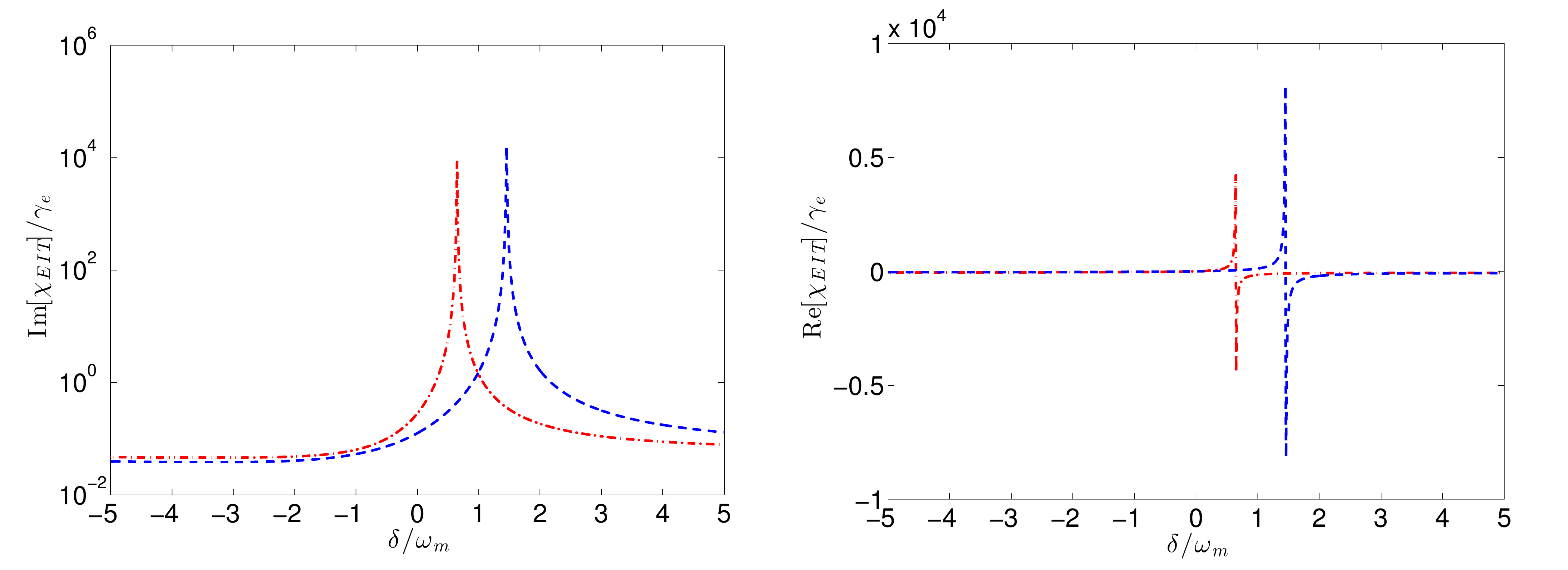}
\caption{(Colors online). Atomic susceptibility for EIT. We consider a sample of atomic $^{87}$Rb in the $|F=1\rangle$ ground state manifold with $\Delta_a = 500\gamma_e$ and $N = 10^{8}$. The control amplitude is $\Omega = 4 \gamma_e$ (red dot-dashed line) and $\Omega = 6 \gamma_e$ (blue dashed line). For the single atom Rabi frequency of the coupling field we assume $\mathcal{E}_a = 2\pi \times 100$ kHz, while the mechanical frequency used only as normalization is $\omega_m = 2\pi \times 300$ kHz.}
\label{fig:Zchi}
\end{center}
\end{figure*}

In terms of the atomic operators $\hat{\sigma}_{ab} = \ket{a}\bra{b}$ the atomic Hamiltonian is
\begin{equation}
H_{\rm atom} = H_0 + H_\mathrm{af},
\end{equation}
where $H_0 = \sum_{a = \{g,e,m\}} \hbar \omega_a \hat{\sigma}_{aa}$ describes the non-interacting internal level structure and the second term contains the interaction of the atomic levels with the control and coupling fields,
\begin{eqnarray} 
H_\mathrm{af} &=& \hbar \Omega \left[e^{i\omega_1t}\hat{\sigma}_{me} +  e^{-i\omega_1t}\hat{\sigma}_{em} \right] \nonumber \\
&+&\hbar\mathcal{E}_a\left[\hat{a}^\dagger e^{i\omega_2t} \hat{\sigma}_{ge} + \hat{a} e^{-i\omega_2t}\hat{\sigma}_{eg}\right].
\end{eqnarray}
Here $\Omega$ is the Rabi frequency for the (classical) control field and $\mathcal{E}_a$ is the real single atom Rabi frequency for the intra-cavity field, $\hat{a}$. We adopt the interaction picture for the atomic levels and work in a rotating frame at the coupling frequency $\omega_2$.  Assuming that the light and the atomic system correlation functions factorize, the equations of motion for the expectation values of the atom-light system in the cavity are then
\begin{align}
&\langle\dot{\hat{a}}\rangle = \left(i \Delta_\mathrm{ca} - \frac{\kappa_\mathrm{ca}}{2}\right)\langle\hat{a}\rangle +\eta_a - i \mathcal{E}_a\langle \hat{\sigma}_{ge}\rangle e^{i \Delta_a t}, \\
&\langle\dot{\hat{\sigma}}_{ge}\rangle =  i \mathcal{E}_a\langle \hat{a}\rangle(\langle\hat{\sigma}_{ee}\rangle -\langle \hat{\sigma}_{gg}\rangle) e^{-i \Delta_a t} - i \Omega\langle \hat{\sigma}_{gm}\rangle e^{-i \Delta_c t}\nonumber\\
&\qquad\quad - \frac{\gamma_e}{2}\langle \hat{\sigma}_{ge}\rangle, \\
&\langle\dot{\hat{\sigma}}_{gm}\rangle =  i \mathcal{E}_a\langle \hat{a}\rangle\langle \hat{\sigma}_{em}\rangle e^{-i \Delta_a t} - i \Omega\langle \hat{\sigma}_{ge}\rangle e^{i \Delta_c t} - \frac{\gamma_m}{2}\langle \hat{\sigma}_{gm}\rangle, \\
&\langle\dot{\hat{\sigma}}_{em}\rangle =  i \mathcal{E}_a\langle \hat{a}^{\dagger}\rangle\langle \hat{\sigma}_{gm}\rangle e^{i \Delta_a t} + i \Omega (\langle\hat{\sigma}_{mm}\rangle - \langle\hat{\sigma}_{ee}\rangle) e^{-i \Delta_c t}\nonumber\\
&\qquad\quad - \frac{\gamma_e}{2} \langle\hat{\sigma}_{em}\rangle.
\end{align}

The terms $\gamma_l$, $l = \{e,m\}$ lead to decay of coherence due to atomic dephasing mechanisms, mainly spontaneous emission from the excited or metastable levels $\ket{e}$, $\ket{m}$. We have also introduced the detunings $\Delta_\mathrm{ca}=\omega_2-\omega_\mathrm{ca}$, $\Delta_a=\omega_2-\omega_{eg}$ and $\Delta_c=\omega_1-\omega_{em}$. For $\Delta_{a}$ and $\Delta_{c}$ much larger than the width of the excited state, the atoms are populating only the lower two states, such that $\langle\hat{\sigma}_{ee}\rangle = 0$  and the total number of atoms is $N = N_g + N_m$ with $\langle\hat{\sigma}_{gg}\rangle = N_g$ and $\langle\hat{\sigma}_{mm}\rangle = N_m$. To find the atomic susceptibility we consider the steady state of the atomic coherences in their respective rest frames and neglect higher order terms in the atom-field coupling $\mathcal{E}_a$
\begin{align}
\langle\hat{\sigma}_{em}\rangle =\,& \frac{\Omega N_m}{\Delta_c - i \gamma_e/2}, \\
\langle\hat{\sigma}_{gm}\rangle =\,& \left[ \Omega \langle\hat{\sigma}_{ge}\rangle - \mathcal{E}_a \langle\hat{a}\rangle \langle\hat{\sigma}_{em}\rangle \right] \frac{1}{\delta - i \gamma_m/2}, \\
\langle\hat{\sigma}_{ge}\rangle =\,& \frac{i \chi_{\rm EIT} \langle\hat{a}\rangle}{-i\mathcal{E}_a}, \\
\chi_{\rm EIT} = \,&\mathcal{E}_a^2 \left(N_g - \frac{\Omega^2 N_m}{(\Delta_c - i \gamma_e/2)(\delta+ i \gamma_m/2)}\right) \nonumber \\
&\times \left(\Delta_a + i\frac{\gamma_e}{2} - \frac{\Omega^2}{\delta + i \gamma_m/2}\right)^{-1}. \label{eq:chiZ}
\end{align}
Here we have introduced the two-photon detuning $\delta = \Delta_a - \Delta_c$. The first term in the atomic susceptibility \eq{chiZ}, dependent on $N_g$, is the usual EIT susceptibility with zero population in the metastable state. In the following we consider that situation and choose $N_g = N$. The equation of motion for the expectation value of the intracavity field becomes then
\begin{equation}
\frac{d \langle\hat{a}\rangle}{dt} =  \left(i \Delta_\mathrm{ca} -\frac{ \kappa_\mathrm{ca}}{2}\right) \langle\hat{a}\rangle + \eta_a +  i \chi_{\rm EIT} \langle\hat{a}\rangle,
\label{eq:fieldEOMz}
\end{equation}
and the steady state for the cavity field is 
\begin{equation}
 \langle\hat{a}\rangle = \frac{\eta_a}{-i \Delta_\mathrm{ca} + \kappa_\mathrm{ca}/2 -  i \chi_{\rm EIT}}, 
\label{eq:ssfieldZ}
\end{equation}
which clearly shows the modification of the cavity response by the the atomic system in the form of a modified detuning $\Delta_\mathrm{af} = \Delta_\mathrm{ca} + \mathrm{Re}[\chi_{\rm EIT}]$ and decay rate $\kappa_\mathrm{af} = \kappa_\mathrm{ca} + \mathrm{Im}[\chi_{\rm EIT}]$ with 
\begin{widetext}
\begin{eqnarray}
\mathrm{Re}[\chi_{\rm EIT}] &=& - \mathcal{E}_a^2 N \left[\Delta_a \left(\delta^2 + \frac{\gamma_m^2}{4}\right) - \Omega^2 \delta \right] \left[\left(\Delta_a^2 + \frac{\gamma_e^2}{4}\right)\left(\delta^2 + \frac{\gamma_m^2}{4}\right) + \Omega^4 - 2\Omega^2 \left(\Delta_a\delta - \frac{\gamma_e\gamma_m}{4} \right) \right]^{-1}, \\
\mathrm{Im}[\chi_{\rm EIT}] &=& \mathcal{E}_a^2 N \left[\frac{\gamma_e}{2} \left(\delta^2 + \frac{\gamma_m^2}{4}\right) + \Omega^2 \frac{\gamma_m}{2}\right]\left[\left(\Delta_a^2 + \frac{\gamma_e^2}{4}\right)\left(\delta^2 + \frac{\gamma_m^2}{4}\right) + \Omega^4 - 2\Omega^2 \left(\Delta_a\delta - \frac{\gamma_e\gamma_m}{4} \right) \right]^{-1}. 
\end{eqnarray}
\end{widetext}

Figure \ref{fig:Zchi} shows the imaginary and real parts of the EIT susceptibility as a function of the two-photon detuning.  It displays are two main features: an \emph{atomic resonance},  due to the metastable state $|m\rangle$ dressed by the strong control field; and the \emph{EIT} or \emph{two-photon resonance}, due to the Raman coherence established between $|g\rangle$ and $|m\rangle$. The latter is located at $\delta = 0$ and it is characterized by a vanishing dispersion, while the former is dominated by atomic absorption and its position is set by the dynamical Stark shift, $\delta = \Omega^2/\Delta_a$. The strength of the absorption is directly proportional to the total number of atoms and inversely proportional to the single-photon detuning, since the coupling is mediated by the excited state, and its linewidth may by tuned by changing the amplitude $\Omega$.
 
\begin{figure}[htbp]
\begin{center}
\includegraphics[width = \columnwidth]{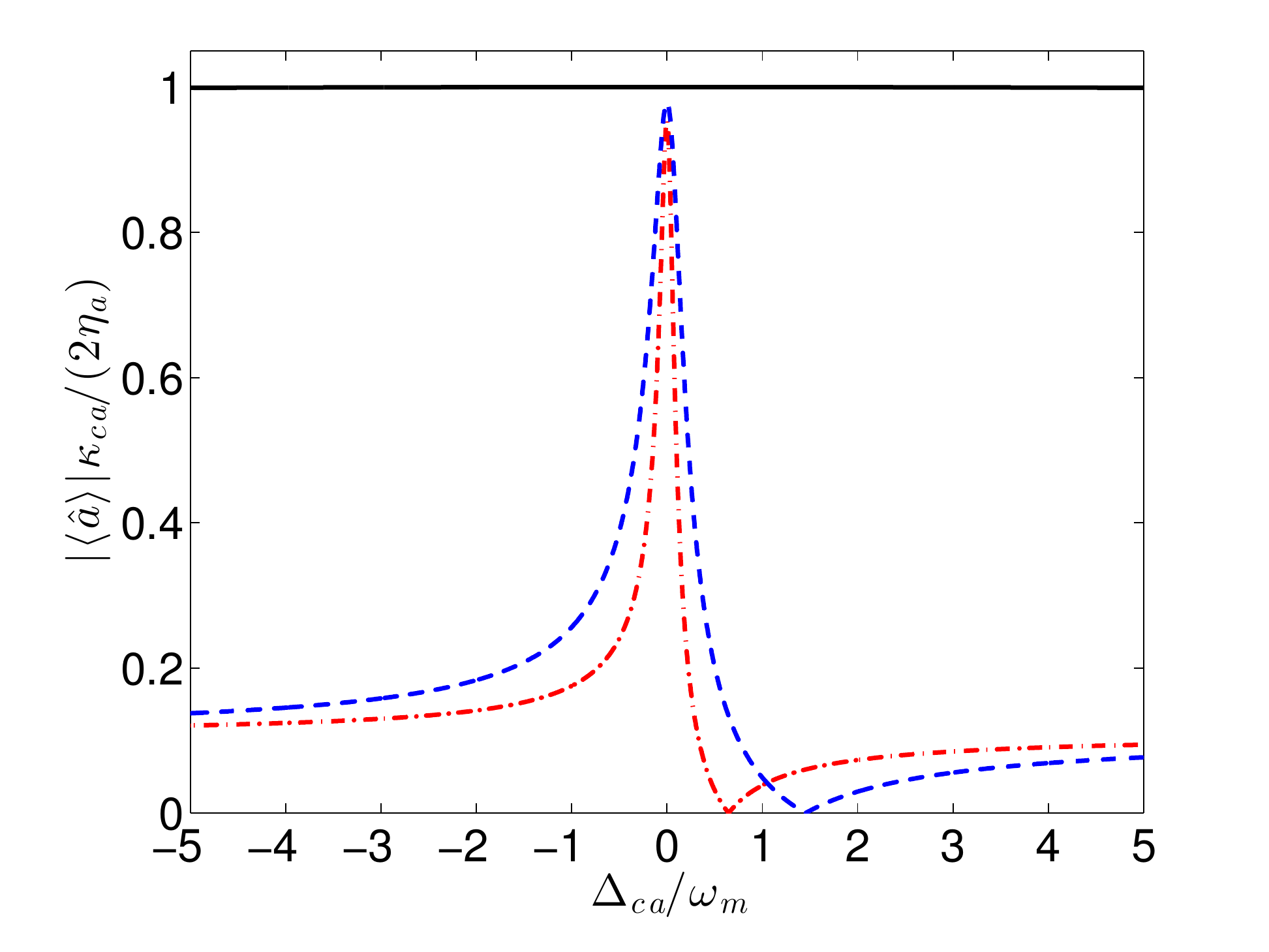}
\caption{(Colors online). Normalized intracavity electric field of the atomic cavity. The cavity has a decay rate $\kappa_{ca} = 2\pi \times 70$~MHz and it is resonant with the two-photon resonance of EIT, $\delta = \Delta_{ca}$. We consider an atomic ensemble in EIT configuration with $N = 10^8$ atoms and control amplitude: $\Omega = 4 \gamma_e$ (red dot-dashed line), $\Omega = 6 \gamma_e$ (blue dashed line). The black solid line represents the cavity response without atoms.}
\label{fig:alpha}
\end{center}
\end{figure}

Figure \ref{fig:alpha} shows the corresponding intracavity field amplitude as a function of the detuning from the \emph{cavity resonance}. Note the drastic reduction in the effective cavity line width as a result of the interaction with the atoms. The origin of this narrowing is the interplay between atomic absorption and dispersion. There is a strong suppression corresponding to the atomic resonance, where the light is almost completely absorbed by the atomic cloud. The atomic dispersion corresponding to the real part of the susceptibility has a smaller peak value than its imaginary counterpart, but its baseline tends to a constant value away from the resonance and it affects the cavity response. Note that the position of the cavity resonance is not shifted as compared to the bare case since the atomic dispersion vanishes at the two-photon resonance, a well-known property of EIT. The absorption at cavity resonance determines the small difference between the bare cavity response and the atomic case. 
The effective linewidth of the cavity is also influenced by the separation between the atomic and EIT resonances: this can be controlled via the intensity of the control field $\Omega$, as shown in \fig{alpha}, or via the single-photon detuning $\Delta_a$.

\subsection{Recoil Induced Resonances}
We turn next to an atomic $\Lambda$ system that relies on the motional states of the ultracold gas. In particular, we evaluate the optical response of the gas in the vicinity of a recoil-induced resonance wherein the motion of the atoms under the influence of optical fields can mediate the conversion of atomic kinetic energy into radiation. We consider an atomic cloud of $N$ atoms confined in a cylindrically symmetric trap with the axial confinement significantly weaker than the radial confinement. This one-dimensional geometry enhances atomic recoil effects along the weakly confined axis. Along this axis, the atoms are illuminated by a strong control laser with frequency $\omega_1$ and wave vector $k_1$, and a weak counter-propagating coupling laser with frequency $\omega_2$ and wave vector $k_2$. The relevant energy levels are sketched in Fig. \ref{fig:EITschemes}C. The energy difference between the two lower lying states can be changed by detuning one of the lasers. 

In the absence of collisions but accounting for photon recoil, the Hamiltonian describing the interaction of the atomic ensemble with the light fields is $H=\sum_k H_k$, with
\begin{eqnarray}\nonumber
H_k&=&\frac{\hbar^2k^2}{2m_a}\hat{c}_g^\dagger(k)\hat{c}_g(k)+\left(\frac{\hbar^2k^2}{2m_a}+\hbar\omega_0\right)\hat{c}_e^\dagger(k)\hat{c}_e(k)\\ \nonumber
&+&\hbar\Omega\left[i e^{i\omega_1t}\hat{c}_g^\dagger(k-k_1)\hat{c}_e(k) + \mathrm{h.c.} \right] \\
&+&\hbar\left[i \mathcal{E}_a\hat{a}^\dagger e^{i\omega_2t}\hat{c}_g^\dagger(k-k_2)\hat{c}_e(k) + \mathrm{h.c.}\right] 
\label{eq:aHamiltonian}
\end{eqnarray}
Here $\hat{c}_{g}(k)$ and $\hat{c}_{e}(k)$ are the annihilation operator of a ground  and excited state atom with momentum $\hbar k$, respectively. They follow standard bosonic commutation relations, $\left[\hat{c}_i(k),\hat{c}_j^\dagger(k')\right]=\delta_{kk'}\delta_{ij}$. As before the Rabi frequency for the control field is $\Omega$ and $\mathcal{E}_a$ is the single atom Rabi frequency for the intracavity field, $\hat{a}$. 

The expectation value for the intracavity field is governed by the equation of motion
\begin{equation}\label{eq:fieldeom1}
\langle\dot{\hat{a}}\rangle=-\left(i\omega_\mathrm{ca}+ \frac{\kappa_\mathrm{ca}}{2}\right)\langle\hat{a}\rangle+\mathcal{E}_a e^{i\omega_2 t}\sum_k\langle \hat{c}_g^\dagger(k-k_2)\hat{c}_e(k)\rangle+\eta_a,
\end{equation}
where we have added cavity dissipation and drive in the familiar fashion.

We introduce the single-photon detuning $\Delta_a=\omega_0-\omega_1$, and the two-photon detuning, $\delta=\omega_2-\omega_1$ and, as in the case for EIT, assume a large single-photon detuning such that $(\delta,\Omega) \ll \Delta_a$. In this limit, we can adiabatically eliminate the excited state and evaluate the equations of motion for the ground state populations and momentum state coherences of the atomic gas \cite{Hafezi2008}. Details of this calculation are outlined in Appendix A. 
  
We define the momentum-dependent population in the ground state, $\Pi_p$, and the ground state coherences between adjacent momentum classes, $\zeta_{p \pm 1}$, as 
\begin{widetext}
\begin{eqnarray}
 \Pi_p&=&\rho_{gg}(k,k) =\langle \hat{c}_g^\dagger(k)\hat{c}_g(k)\rangle= N_g,\nonumber \\
 \zeta_{p+1}&=&\rho_{gg}(k+2k_0,k)e^{-i\delta t}=\rho_{gg}(p+1,p)e^{-i\delta t}=\langle \hat{c}_g^\dagger(k)\hat{c}_g(k+2k_0)\rangle e^{-i\delta t},\nonumber \\
  \zeta_{p-1}&=&\rho_{gg}(k-2k_0,k)e^{i\delta t}=\rho_{gg}(p-1,p)e^{i\delta t}=\langle \hat{c}_g^\dagger(k)\hat{c}_g(k-2k_0)\rangle e^{i\delta t},
  \end{eqnarray}
 Assuming an initial ground state population momentum distribution $\Pi_{\rm th}$ that is in thermal equilibrium at temperature $T_a$, we can write the coupled system of equations for the atomic populations and coherences as
 \begin{eqnarray}
 \frac{d}{dt}\Pi_p&=&i\beta\mathcal{E}_a\langle\hat{a}\rangle\left(\zeta_{p+1}-\zeta_{p-1}\right)+i\beta \mathcal{E}_a \langle\hat{a}^{\dagger}\rangle\left(\zeta_{p-1}-\zeta_{p+1}^*\right) -\gamma_{\rm pop}\Pi_p+\gamma_{\rm pop}\Pi_{{\rm th},p}\  \label{eq:gseom1}\\
 \frac{d}{dt}\zeta_{p+1}&=&-\left(4i\omega_r(2p+1)+i\delta+\gamma_{\rm coh}\right)\zeta_{p+1} -i\beta \mathcal{E}_a \langle\hat{a}^{\dagger}\rangle(\Pi_{p+1}-\Pi_{p}),\\ 
 \frac{d}{dt}\zeta_{p-1}&=&\left(4i\omega_r(2p-1)+i\delta-\gamma_{\rm coh}\right)\zeta_{p-1} -i\beta \mathcal{E}_a\langle\hat{a}\rangle(\Pi_{p-1}-\Pi_{p}),
 \end{eqnarray}
 \end{widetext}
 where 
 \begin{equation}
 \omega_r=\hbar k^2/2m_a
 \end{equation}
 is the atomic recoil frequency,  $2k_0=k_1-k_2$, the dimensionless momentum $p=\hbar k/(2\hbar k_0)$, and we have introduced the normalized control strength $  \beta=\Omega/\Delta_a$.
The last two terms in the equation for the population are due to the fluctuation-dissipation theorem.  

\begin{figure*}[hbt]
\begin{center}
\includegraphics[width=\textwidth]{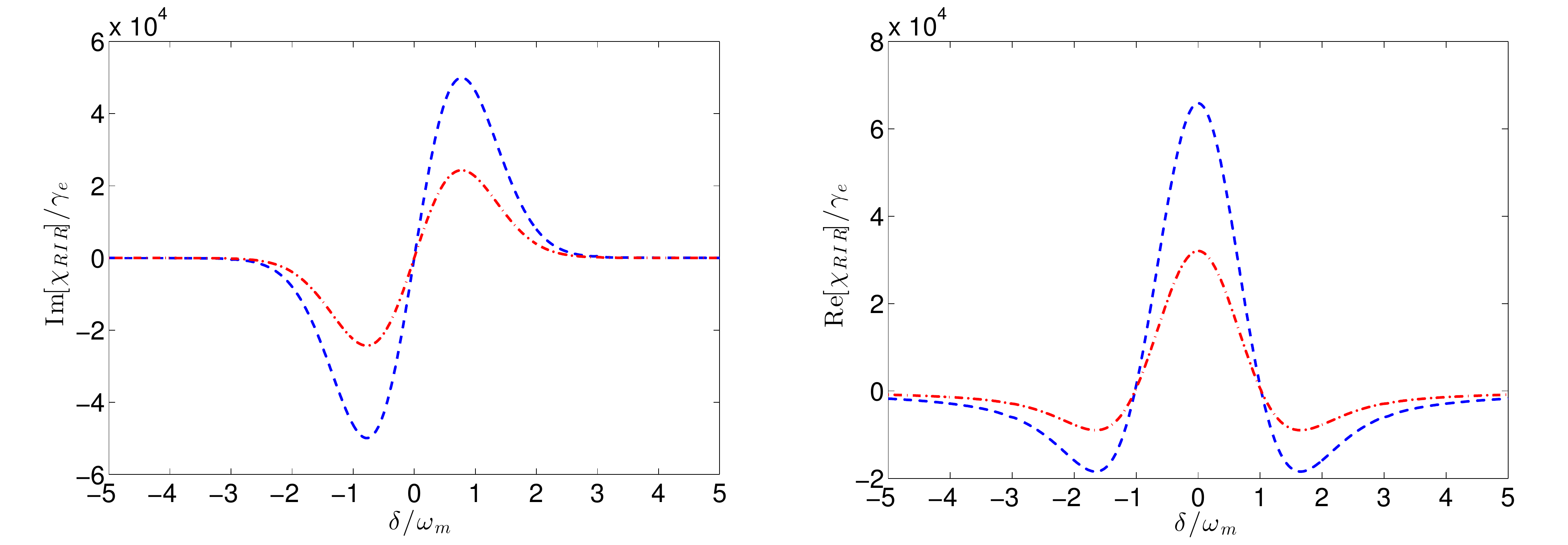}
\caption{ (Color online) Real and imaginary parts of the atomic susceptibility for a thermal ensemble of atoms at [$T_a=21 \mu$K, $\Omega= 1.8 \gamma_e$] (red dot-dashed line),  [$T_a= 21\mu$K, $\Omega = 2.6 \gamma_e$] (blue dashed line). Here $N = 10^8$ atoms, $\Delta_a=-15  \gamma_{e}$, $\omega_r=2 \pi \times 3.77$~kHz, $\gamma_{e}=2\pi \times 6.07 $~MHz,  $\gamma_{\rm coh}=2 \pi \times 10$~kHz, and $\mathcal{E}_a= 2 \pi \times 500$~kHz. }
\label{fig:RIRchi}
\end{center}
\end{figure*}

Assuming that the atomic populations remain in thermal equilibrium and that the coherences reach steady state over the time scale of the evolution of the electric fields,  the equation of motion for the mean intracavity field becomes 
 \begin{equation}
  \frac{d \langle\hat{a}\rangle}{dt}=\left(i\Delta_\mathrm{ca}-\frac{\kappa_\mathrm{ca}}{2}\right)\langle\hat{a}\rangle+\eta_a+i\chi_{\rm RIR} \langle\hat{a}\rangle,
  \label{eq:fieldEOMr}
\end{equation}
where $\Delta_\mathrm{ca}=\omega_2-\omega_\mathrm{ca}$ is the detuning of the cavity from the coupling frequency. This is essentially the same as for EIT, except that the susceptibility is now  $\chi_{\rm RIR}$. For a strong control beam that does not suffer any significant depletion, we can solve for the steady state of the intra-cavity field to get
\begin{equation}
\langle\hat{a}\rangle = \frac{\eta_a}{-i \Delta_\mathrm{ca} + \kappa_\mathrm{ca}/2 -  i \chi_{\rm RIR}}. 
\label{eq:ssfieldRIR}
\end{equation} 
with
\begin{widetext}
\begin{eqnarray}\nonumber
{\rm Re}[\chi_{\rm RIR}]&=& \frac{\mathcal{E}_a ^2N}{\Delta_a}+ (\beta \mathcal{E}_a)^2N\left\{ \sum_p\frac{\Pi_{{\rm th},p}(\delta+4\omega_r(2p+1))}{(\gamma_{\rm coh}^2+(\delta+4\omega_r(2p+1))^2)}-\frac{\Pi_{{\rm th},p}(\delta+4\omega_r(2p-1))}{(\gamma_{\rm coh}^2+(\delta+4\omega_r(2p-1))^2)}\right\},\\
 {\rm Im}[\chi_{\rm RIR}]&=&- (\beta \mathcal{E}_a)^2N\gamma_{\rm coh}\left\{\sum_p\frac{\Pi_{{\rm th},p}}{(\gamma_{\rm coh}^2+(\delta+4\omega_r(2p+1))^2)}-\frac{\Pi_{{\rm th},p}}{(\gamma_{\rm coh}^2+(\delta+4\omega_r(2p-1))^2)}\right\}.
 \label{eq:chi}
\end{eqnarray}
\end{widetext}

Figure \ref{fig:RIRchi} shows ${\rm Im}[\chi_{\rm RIR}]$ and ${\rm Re}[\chi_{\rm RIR}]$  as functions of the two-photon detuning for a thermal ensemble of ultracold atoms. The atomic susceptibility depends on the number of atoms $N$, the Rabi frequency $\Omega$, the single-photon detuning $\Delta_a$ and the temperature of the ensemble $T_a$. The decoherence rate  $\gamma_{\rm coh}$ depends both on off-resonant light scattering as well as atomic collisions. For a laser cooled atomic gas, decoherence rates as low as 1 ms$^{-1}$ have been demonstrated \cite{Hafezi2008}. This is more than two orders of magnitude smaller than the typical mechanical resonance frequency $\omega_m$ that we consider in this work.

As with EIT, the RIR results in modifications to the atomic susceptibility with the detuning changing to $ \Delta_\mathrm{af}=\Delta_\mathrm{ca}+  {\rm Re}[\chi_{\rm RIR}]$ and the decay rate changing to $ \kappa_\mathrm{af}=\kappa_\mathrm{ca}/2+  {\rm Im}[\chi_{\rm RIR}]$. As will become apparent in Section IV, the effect of recoil resonances on the coupling field is formally analogous to the optomechanical effects inside a cavity with a moving mirror, with the difference that instead of having a single frequency as is the case in single-mode optomechanics, we now have a distribution of frequencies associated with the center-of-mass momentum distribution of the atoms. The presence of a negative susceptibility $\rm Im[\chi_{RIR}]$ for negative detunings $\delta$ is indicative of gain in the atomic medium, leading to an exponential build up of the coupling laser in the linear, small signal regime. 

\begin{figure}[htb]
\begin{center}
\includegraphics[width=\columnwidth]{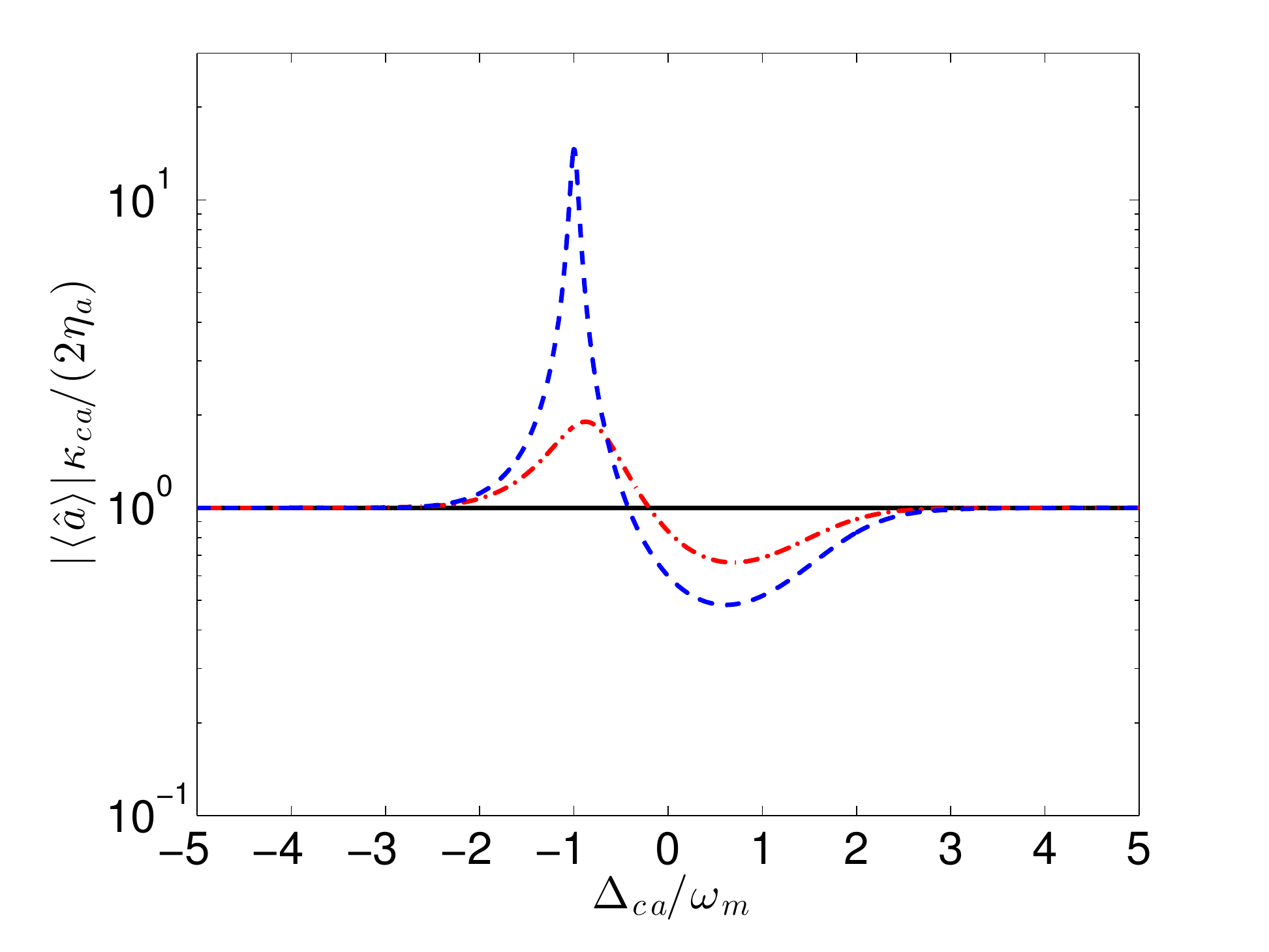}
\caption{(Color online). Normalized steady state coupling field after the atomic medium for the parameters of Fig.~\ref{fig:RIRchi}. 
The atomic ensemble is in free space which corresponds to a decay rate $\kappa_{ca} = c/L_a = 2\pi \times 600$ GHz, and with $\Delta_\mathrm{ca}=0$.
We show the result for [$T_a=21 \mu$K, $\Omega= 1.8 \gamma_e$] (red dot-dashed line),  [$T_a= 21\mu$K, $\Omega = 2.6 \gamma_e$] (blue dashed line) and without atomic medium (black solid line).}
\label{fig:RIRprobe}
\end{center}
\end{figure}

Figure \ref{fig:RIRprobe} shows the normalized cavity field amplitude as a function of the cavity detuning $\Delta_{ca}$ for the parameters of Fig.~\ref{fig:RIRchi}. The field amplitude is strongly suppressed for small positive detunings. Also, for various combinations of scaled control fields $\beta$ and temperature $T_a$, there is a dramatic build-up of intensity for a narrow range of frequencies at negative detuning. Within this window, the atomic gas mediates the coherent transfer of energy from the control field to the coupling field, leading to gain in the latter. The gain feature can be tuned in frequency by varying the control detuning, intensity and the temperature of the atomic ensemble. 

Importantly we note that while the linearized theory would predict an exponential growth of the coupling field with increasing atom number, the actual gain is limited in practice by depletion of the control field. Such saturation effects are not accounted for in the present description. Our specific examples  of experimental parameters have been chosen so as to safely stay away from such limiting effects. 

Summarizing this section, both the EIT and RIR schemes lead to frequency dependent atomic susceptibilities with spectrally narrow features as shown in  Fig.~\ref{fig:Zchi} and \fig{RIRchi}. In the EIT scheme, the two-photon resonance corresponds to a narrow window of vanishing absorption and dispersion that leaves the coupling field unchanged. In the case of RIR, the atoms act as a gain medium, converting control photons to coupling field photons via atomic recoil, leading to an enhancement of the latter within a narrow range of frequencies as shown in Fig. \ref{fig:RIRprobe}. Both effects can be used to enhance optomechanical cooling.  In the EIT scheme, this results from a reduction of the effective linewidth of the cavity down to the range of the transparency window of the atomic medium while in the RIR scheme, the atomic gas acts as a gain medium enhancing the coupling field around the anti-Stokes sideband. 

\section{Hybrid atom-optomechanics}
We are now in a position to investigate the effect of the atomic ensemble on the cooling properties of the hybrid optomechanical setups of section II.  From the Hamiltonian~(\ref{eq:totalH}) the equations of motion for the optomechanical cavity field $\hat{c}$ and the mechanical mode $\hat{b}$ are
\begin{eqnarray}
\dot{\hat{c}}&=&-\left(i\omega_\mathrm{cm}+\frac{\kappa_\mathrm{cm}}{2}\right)\hat{c}+\eta_c-iJ\hat{a}-ig_0\hat{c}(\hat{b}+\hat{b}^\dagger),\\
\dot{\hat{b}}&=&-\left(i\omega_{m}+\frac{\gamma_{m}}{2}\right)\hat{b}-ig_0\hat{c}^\dagger\hat{c},
\end{eqnarray}
where we have introduced the mechanical damping rate $\gamma_m = \omega_m/Q$, with $Q$ the quality factor of the mechanics. We assume $\kappa_\mathrm{ca} \gg \kappa_\mathrm{cm}$ which implies that, over the time-scale of the dynamics of the optomechanical system, the atomic cavity follows adiabatically the dynamics of the mechanical resonator. 

\begin{figure}[h]
\begin{center}
\includegraphics[width=\columnwidth]{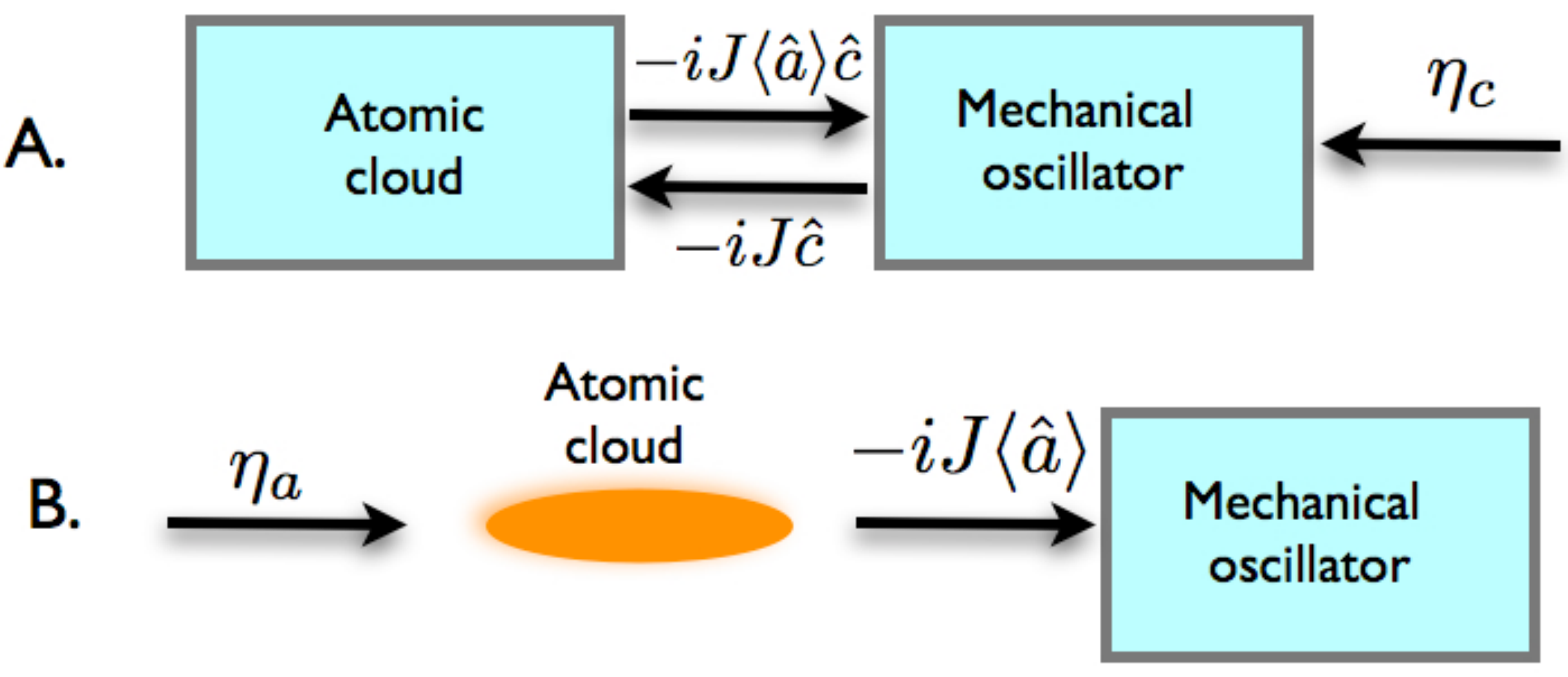}
\caption{(Colors online). Schematic of the two atom mediated optomechanical coupling schemes. In setup A, the atomic cavity provides a feedback system for the optomechanical cavity. In setup B, the external drive is filtered through the atomic medium before being injected into the optomechanical cavity.}
\label{fig:coupling}
\end{center}
\end{figure}

We consider in the following the two specific scenarios of  feedback and cascade couplings illustrated in Fig.~\ref{fig:coupling}. In the former case, light is pumped (from the left) into the optomechanical cavity, and then coupled (from the right) into the atomic cavity. In the latter configuration, the driving field first propagates through the atomic medium and is then injected into the optomechanical resonator.

\subsection{Feedback coupling}
Consider first the feedback scheme of Fig.~\ref{fig:coupling}A. The output of the optomechanical cavity drives the atomic cavity, so that the cavity field driving term $\eta_a$ in Eqs.~(\ref{eq:ssfieldZ}) and (\ref{eq:ssfieldRIR}) is now $-iJ\hat{c}$. The assumption that the atomic cavity follows adiabatically the evolution of the optomechanical cavity field allows to replace $\langle \hat{a} \rangle$ by 
\begin{equation}
\langle\hat{a}\rangle = \frac{iJ\langle\hat{c}\rangle}{i (\Delta_\mathrm{ca}  +\chi)- \kappa_\mathrm{ca}/2},
\end{equation}
where we dropped the subscript for the atomic susceptibility for notational convenience. In a frame rotating at the drive frequency $\omega_2$, the equation of motion for $\langle \hat{c} \rangle $ then simplifies to
\begin{eqnarray}
\langle \dot{\hat c}\rangle &=&i\left(\Delta_\mathrm{cm}-\frac{\kappa_{\rm cm}}{2}- g_0\langle\hat{b}+\hat{b}^\dagger\rangle\right)\langle\hat{c}\rangle -iJ\langle \hat a \rangle +\eta_c\nonumber \\
&=&i\left(\Delta_\mathrm{cm}- g_0\langle\hat{b}+\hat{b}^\dagger\rangle\right)\langle\hat{c}\rangle \nonumber\\
&+&\left(-\frac{\kappa_\mathrm{cm}}{2}+\frac{J^2}{i (\Delta_\mathrm{ca}  +\chi)-\kappa_\mathrm{ca}/2}\right)\langle\hat{c}\rangle+\eta_c,
\end{eqnarray}
where $\Delta_\mathrm{cm} = \omega_2 - \omega_\mathrm{cm}$.
It is clear from this expression that the linewidth $\kappa_{cm}$ of the optomechanical cavity is modified by the field in the atomic cavity: this is exactly the effect of the feedback coupling.

\begin{figure*}[ht]
\begin{center}
\includegraphics[width =\textwidth]{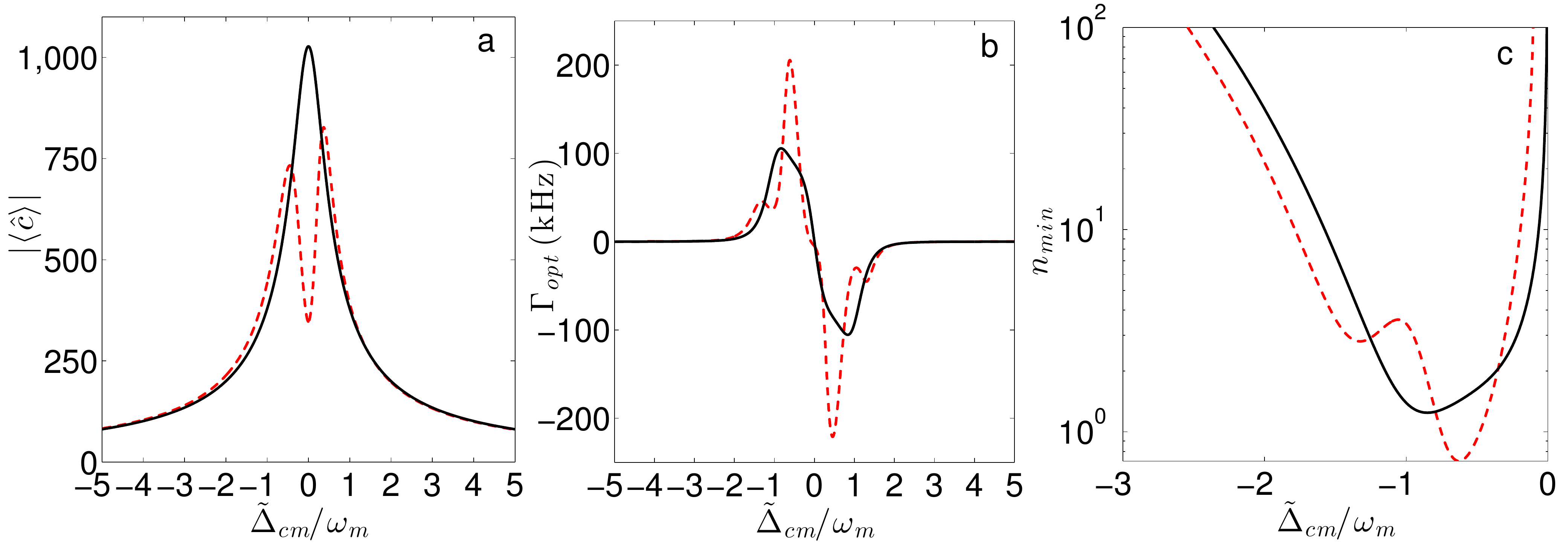}\\
\includegraphics[width =\textwidth]{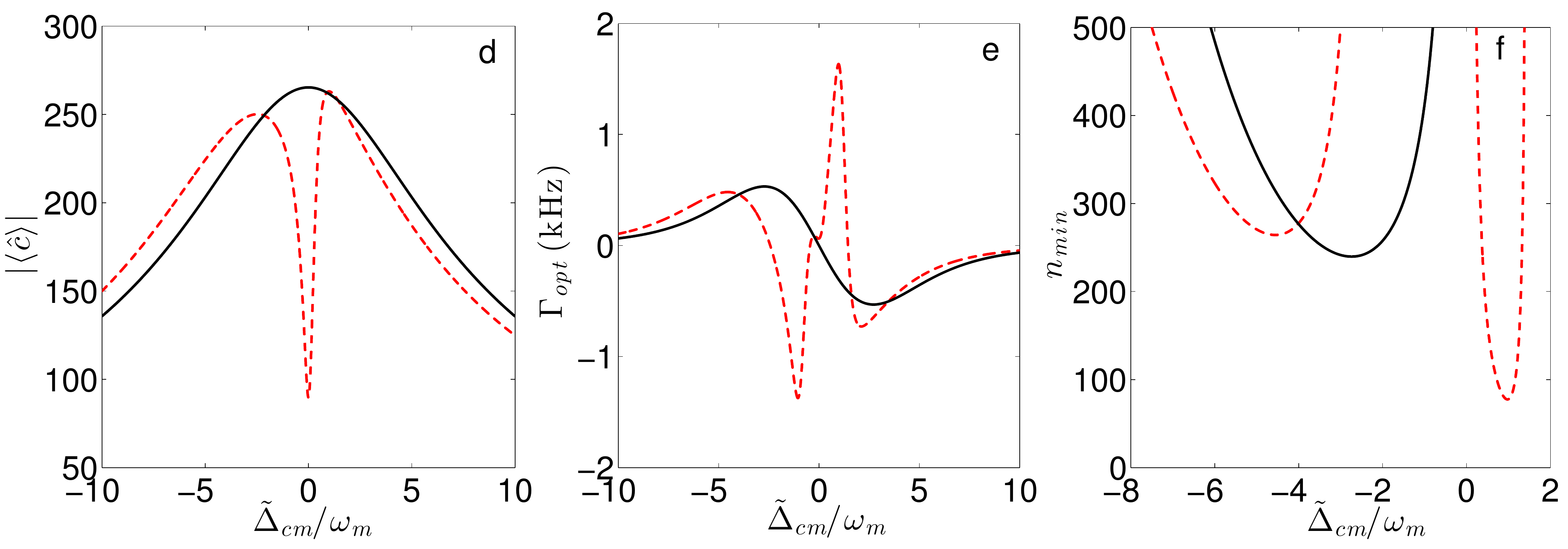}
\caption{(Color online). Cooling characteristics of the hybrid atomic EIT optomechanical system with feedback coupling for  $\omega_m = 2\pi\times 300$~kHz,  $Q = 5 \times 10^7$, $T_{\rm bath} = 300$~K, $P_{\rm in} = 200$~nW, $g_0=2 \pi \times 200$~Hz,  $\kappa_{\rm ca} = 2\pi \times 70$~MHz, $N=10^8$ atoms, $\Omega = 6\gamma_e$, and $\Delta_a = 500 \gamma_e$. Plots (a) and (d) show the mean optomechanical  cavity field $\langle \hat c \rangle$, plots (b) and (e) the optical damping $\Gamma_{\rm opt}$ and plots (c) and (f) the minimum number of phonons as a function of $\tilde{\Delta}_{cm}/\omega_m$ near the resolved sideband regime,  $\kappa_{\rm cm} = 2\pi \times 240$~kHz $ < \omega_m$ (a,b,c), and in the Doppler regime, $\kappa_{\rm cm} = 2\pi \times 3.6$~MHz $ > \omega_m$(d,e,f).  The red dashed lines represent the hybrid case and  the black solid lines represent the case without coupling to the atomic  cavity ($J = 0$) for comparison.}
\label{fig:hybrid_EIT}
\end{center}
\end{figure*}

We now introduce the normalized displacement operator $\hat{x}=\hat{b}+\hat{b}^\dagger$, and we decompose the operators $\hat{c}=\langle \hat c \rangle+\delta\hat{c}$ and $\hat{x}=\langle \hat x \rangle+\delta\hat{x}$, into a classical average value corresponding to the steady state, and small fluctuations around it. Linearizing the equations of motion we arrive at
\begin{equation}
\langle \hat c \rangle=\eta_c\left(-i\tilde{\Delta}_\mathrm{cm}-\frac{J^2}{i (\Delta_\mathrm{ca}  +\chi)-\kappa_\mathrm{ca}/2}+\frac{\kappa_\mathrm{cm}}{2}\right)^{-1},
\label{eq:barc}
\end{equation}
where $\tilde{\Delta}_\mathrm{cm}=\Delta_\mathrm{cm}-g_0\langle \hat x \rangle$. 

The fluctuations in the cavity field are governed by the equation of motion
\begin{equation}
\delta \dot{\hat{c}}=\left(i\tilde{\Delta}_\mathrm{cm}+\frac{J^2}{i (\Delta_\mathrm{ca}  +\chi)-\kappa_\mathrm{ca}/2}-\frac{\kappa_\mathrm{cm}}{2}\right)\delta\hat{c}-ig \delta\hat{x}.
\end{equation}
where we have introduced the linearized coupling constant $g = g_0\langle \hat c \rangle$, which can be taken to be real without loss of generality.
This equation can be solved easily in the Fourier domain to get
\begin{align}\nonumber
&\delta \hat c[\omega]=-ig\delta \hat x[\omega] \times \\
&\left(-i(\tilde{\Delta}_\mathrm{cm}+\omega)-\frac{J^2}{i (\Delta_\mathrm{ca}+\omega +\chi[\omega])-\kappa_\mathrm{ca}/2}+\frac{\kappa_\mathrm{cm}}{2}\right)^{-1},
\end{align}
and $\delta \hat c^\dagger[\omega]=(\delta \hat c[-\omega])^\dagger$. The dynamical radiation pressure force at the mechanical frequency is given by $\delta \hat F_{\rm RP}[\omega_m]=-\hbar G (\delta \hat c[\omega_m]+\delta \hat c^\dagger[\omega_m])$, where $G = g/x_{\rm zpt}$ and $x_{\rm zpt} = \sqrt{\hbar/(2m\omega_m)}$ is the zero point motion of the mechanical oscillator. This gives
\begin{equation}
\delta \hat F_{\rm RP}[\omega_m]=i\hbar \frac{g^2}{x_{\rm zpt}} \delta \hat x[\omega_m]\left(\frac{1}{A^{(+)}-i\omega_m}-\frac{1}{A^{(-)*}-i\omega_m}\right),
\end{equation}
where 
\begin{equation}
A^{(\pm)}=-i\tilde{\Delta}_\mathrm{cm}-\frac{J^2}{i (\Delta_\mathrm{ca} \pm \omega  +\chi^{(\pm)})-\kappa_\mathrm{ca}/2}+\frac{\kappa_\mathrm{cm}}{2},
\end{equation}
and $\chi^{(\pm)}$ is the susceptibility evaluated at $\omega_2\pm\omega_m$. The real and imaginary parts of $\langle \delta \hat F_{\rm RP} \rangle$ change the spring constant and damping rate of the mechanical oscillator via dynamical back-action~\cite{OM_review,OM_cooling}, with
\begin{align}
&\Gamma_{\rm opt}=2 g^2 {\rm Re}\left[\frac{1}{A^{(+)}-i\omega_m}-\frac{1}{A^{(-)*}-i\omega_m}\right],\\
&k_{\rm opt}=2 m \omega_m g^2 {\rm Im}\left[\frac{1}{A^{(+)}-i\omega_m}-\frac{1}{A^{(-)*}-i\omega_m}\right].
\end{align}
We recognize the familiar two components deriving from the Stokes (red-) and anti-Stokes (blue-) sidebands. In particular, the optical damping rate can be written as 
\begin{equation}
\Gamma_{\rm opt}=\Gamma_{\rm anti-Stokes}-\Gamma_{\rm Stokes}.
\label{eq:gammaopt}
\end{equation}
Solving for the steady state minimum occupation number for the mechanical mode coupled to a thermal bath, we obtain
\begin{equation}
n_{\min}=\frac{\Gamma_{\rm Stokes}+\gamma_m n_{\rm bath}}{\Gamma_{\rm opt}+\gamma_m}.
\label{eq:nmin}
\end{equation}
Here, $n_{\rm bath} = k_B T_{\rm bath}/\hbar \omega_m \gg 1$,with  $T_{\rm bath}$ the temperature of the thermal bath of the oscillator. 
If we assume $\omega_m = 2\pi \times 300$ kHz with $Q = 5 \times 10^7$ at room temperature \cite{Chakram2014}, we can estimate the minimum optical damping needed to reach ground state cooling as $\Gamma_{\rm opt} > 2\pi \times 125$~kHz.

\begin{figure}[htb]
\begin{center}
\includegraphics[width =\columnwidth]{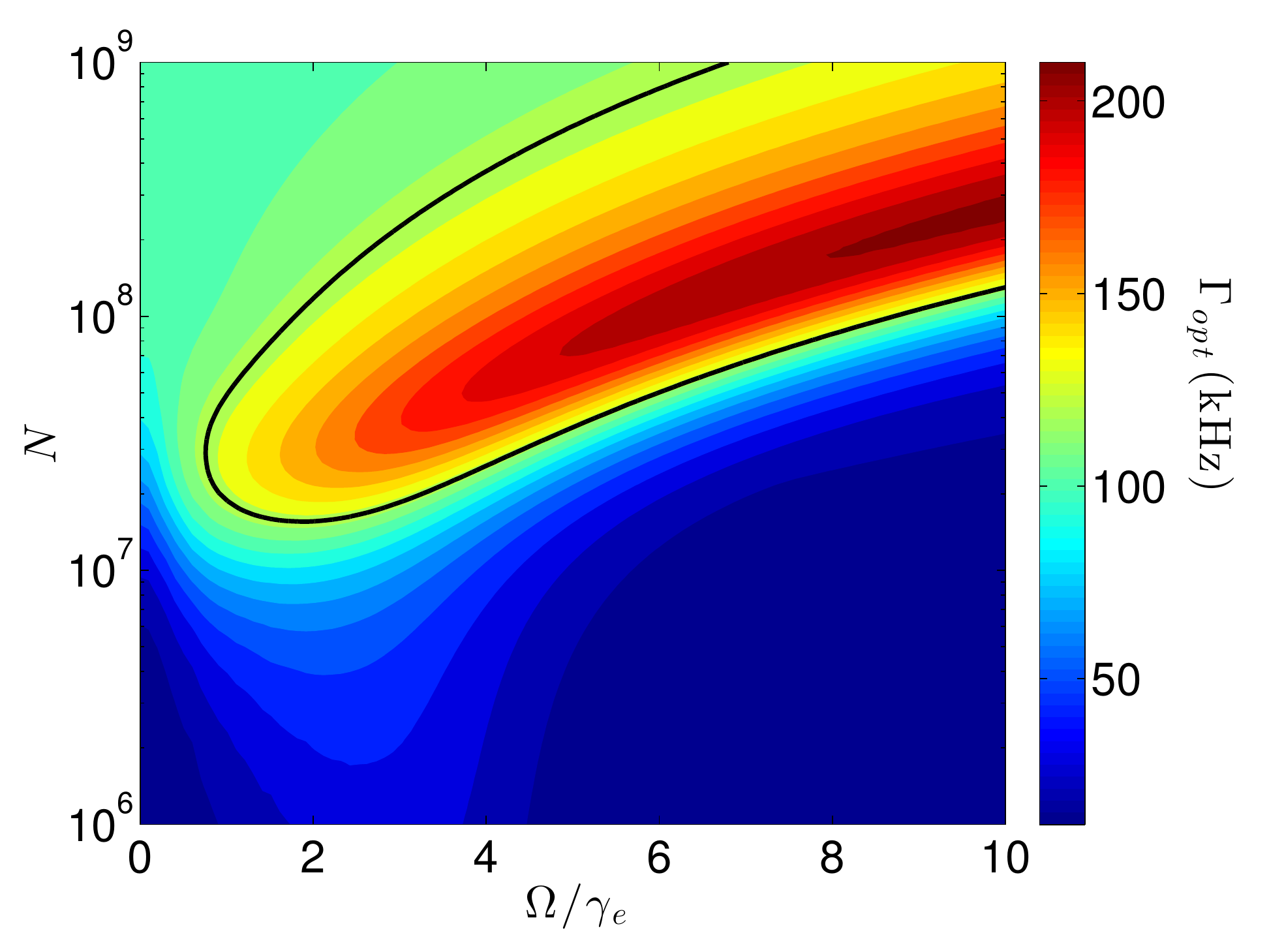}
\caption{(Color online). Maximum optical damping rate for the hybrid atomic EIT optomechanical system with feedback couplingumber $N$. Here $\omega_m = 2\pi\times 300$~kHz with a quality factor $Q = 5 \times 10^7$ and temperature $T_{\rm bath} = 300$~K, $P = 200$~nW, $g_0=2 \pi \times 200$~Hz, $\kappa_{\rm ca} = 2\pi \times 70$~MHz,  $\Delta_a = 500 \gamma_e$, and $\kappa_{\rm cm} = 2\pi \times 240$~kHz. The area enclosed by the black line corresponds to $\Gamma_{\rm opt} > 2\pi \times 125$~kHz and hence to ground state cooling, $n_{\rm min} < 1$, for these parameters. }
\label{fig:EIT_color}
\end{center}
\end{figure}

Figure \ref{fig:hybrid_EIT}  
summarizes important cooling features of the EIT based hybrid optomechanical system and compares them to the purely optomechanical cooling ($J=0$) situation. The upper plots (a,b,c) are for an intermediate  situation close to the resolved side band regime of optomechanics, with $\kappa_{\rm cm} \approx \omega_m$, and the lower series of plots (d,e,f) for the so-called Doppler regime $\kappa_{\rm cm} \gg \omega_m$. 
Remarkably, we find in the intermediate regime a configuration that leads to ground state cooling from room temperature, as clearly shown in Fig.~\ref{fig:hybrid_EIT}(c).  Introducing the hybrid system improvement factor
\begin{equation}
\label{eq:improvf}
\xi = n_{\rm min}^{\rm cm}/n_{\rm min}^{\rm cm + \rm ca},
\end{equation}
we have $\xi \approx 2$ for this situation, a value necessary to obtain $n_{\rm min} < 1$ in that case. We observe also that the best cooling is obtained for $-\omega_m < \tilde{\Delta}_{\rm cm} < 0$ on the red side of the resonance, but slightly shifted compared to the familiar resolved side band condition $\tilde{\Delta}_{\rm cm} = -\omega_m$. 

Figure \ref{fig:hybrid_EIT}(f) shows that in the Doppler regime the improvement factor increases to $\xi \approx 3$, even though the system is cooled to a mean phonon number still far removed from the ground state. Interestingly, though, the strongest cooling feature is now on the blue-side of the resonance, corresponding to $\tilde{\Delta}_{\rm cm} = \omega_m$.

We can gain some degree of intuitive understanding of these results by considering the intracavity fields of Eq.~\eq{barc}, see~\fig{hybrid_EIT}(a,d). As is well known \cite{OM_cooling} the optical damping finds its origin in the difference between the two mechanically generated sidebands, located at $\tilde{\Delta}_{\rm cm} = \omega_2 \pm \omega_m$, whose shape is determined by the intracavity field. Without feedback each sideband has a single peak. Their difference is always positive on the red-side of the resonance, corresponding to cooling. Instead, in the hybrid configuration there is a dip in the field at resonance. The situation is less clear-cut in the presence of feedback, and a more detailed quantitative analysis is required in general to understand the detailed features of cooling, in particular whether it occurs on the red or blue-detuned side of the resonance. First, we note that at the atomic resonance the light entering into the feedback cavity is completely absorbed, thus providing no coupling back to the optomechanical cavity. Second, for situations where  the atomic cavity field is small, the field in the optomechanical cavity remains closer to the uncoupled $(J = 0)$ case. Finally, at the cavity resonances $\tilde{\Delta}_{\rm cm} = \Delta_{\rm ca} = 0$, where the atomic susceptibility vanishes, the feedback field simply contributes an additional term $2J^2/\kappa_\mathrm{\rm ca}$ to the optomechanical cavity linewidth. This induces a dip in the field that is absent without feedback, see dotted red lines in \fig{hybrid_EIT}a and \fig{hybrid_EIT}d.  Away from these limiting situations both the spectral properties and amplitude of the feedback field depend on the linewidth established by the combined effects of $\Omega$ and $N$. In the extreme Doppler regime $\kappa_{\rm cm} \gg \omega_m$, the widths of the sidebands are much broader than their separation, see \fig{hybrid_EIT}d. This results in a situation opposite to the familiar resolved side band regime, with cooling on the blue-side and heating on the red-side, see \fig{hybrid_EIT}(e). When reducing $\kappa_{\rm cm}$ and moving towards the resolved sideband regime, the two peaks in the field start to be be resolved, and this enhances cooling on the red side of the cavity resonance as emerges from \fig{hybrid_EIT}(b). 

Figure \ref{fig:EIT_color} shows the dependence of the optical damping on the control Rabi frequency $\Omega$  and atom number $N$ for the case leading to ground state cooling from room temperature. Importantly, the parameter region that results in such cooling is large, an indication of the robustness of the hybrid system approach with respect to parameter fluctuations.  

\begin{figure}[h]
\begin{center}
\includegraphics[width = 1\columnwidth]{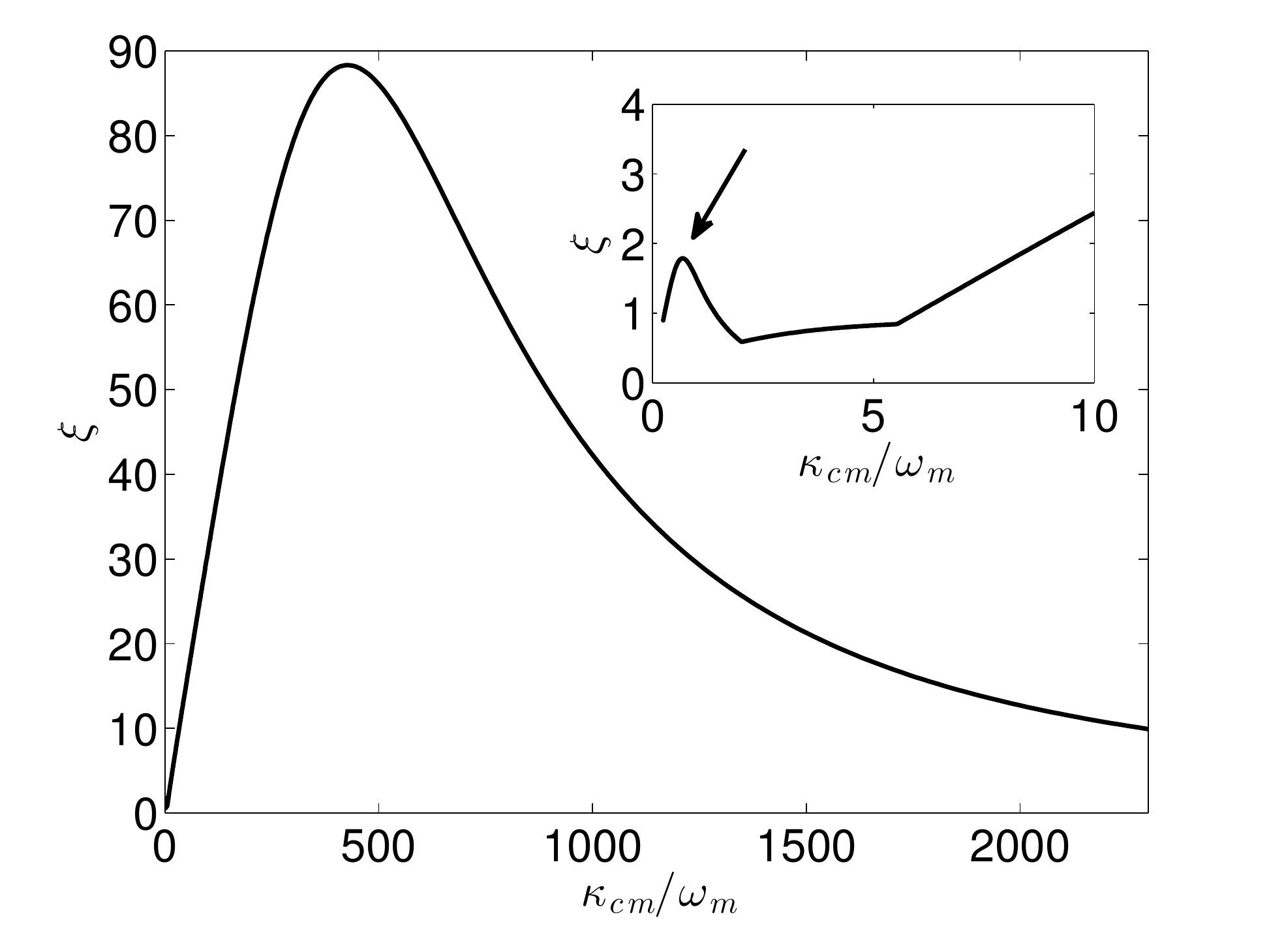}
\caption{Improvement factor $\xi$ when varying the cavity damping rate from the Doppler (right) towards the resolved-sideband (left) regime. The inset shows the area where ground state cooling may be achieved (arrow). Here $\omega_m = 2\pi\times 300$~kHz, $Q = 5 \times 10^7$, $T_{\rm bath} = 300$~K, $P = 200$~nW, $g_0=2 \pi \times 200$~Hz, $\kappa_{ca} = 2\pi \times 70$~MHz, $N=10^8$ atoms, $\Omega = 6\gamma_e$and $\Delta_a = 500 \gamma_e$.}
\label{fig:EIT_improvement}
\end{center}
\end{figure}

Finally, \fig{EIT_improvement} plots the improvement factor $\xi$ for a large range of optomechanical cavity decay rates. Deep in the Doppler regime the atomic ensemble provides an improvement of almost two orders of magnitude over conventional cooling. This  factor decreases as one approaches the resolved side-band regime, but interestingly,  the border between the two regimes is characterized by a feature that allows for ground state cooling as highlighted in the inset. We remark that the advantage of working in an intermediate regime between the resolved side band and Doppler regime has also recently been pointed out in Ref.~\cite{PKU} in a different context.
 
\subsection{Cascade coupling}

\begin{figure*}
\begin{center}
\includegraphics[width=\textwidth]{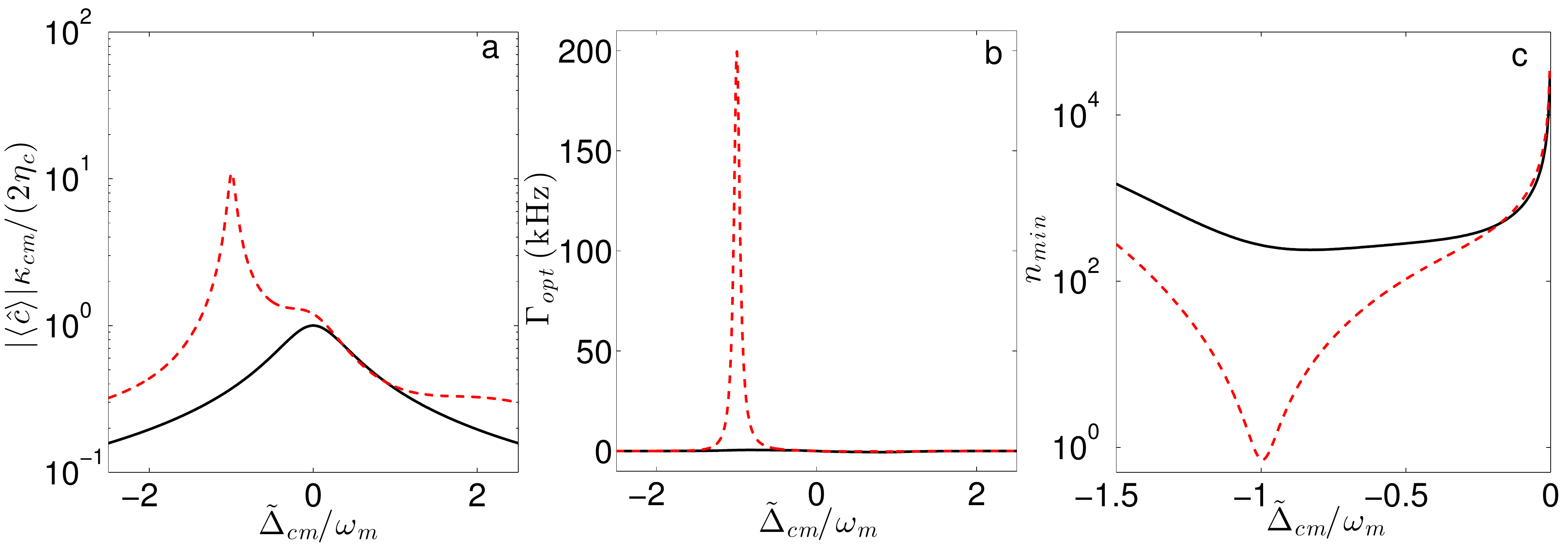}\\
\includegraphics[width=\textwidth]{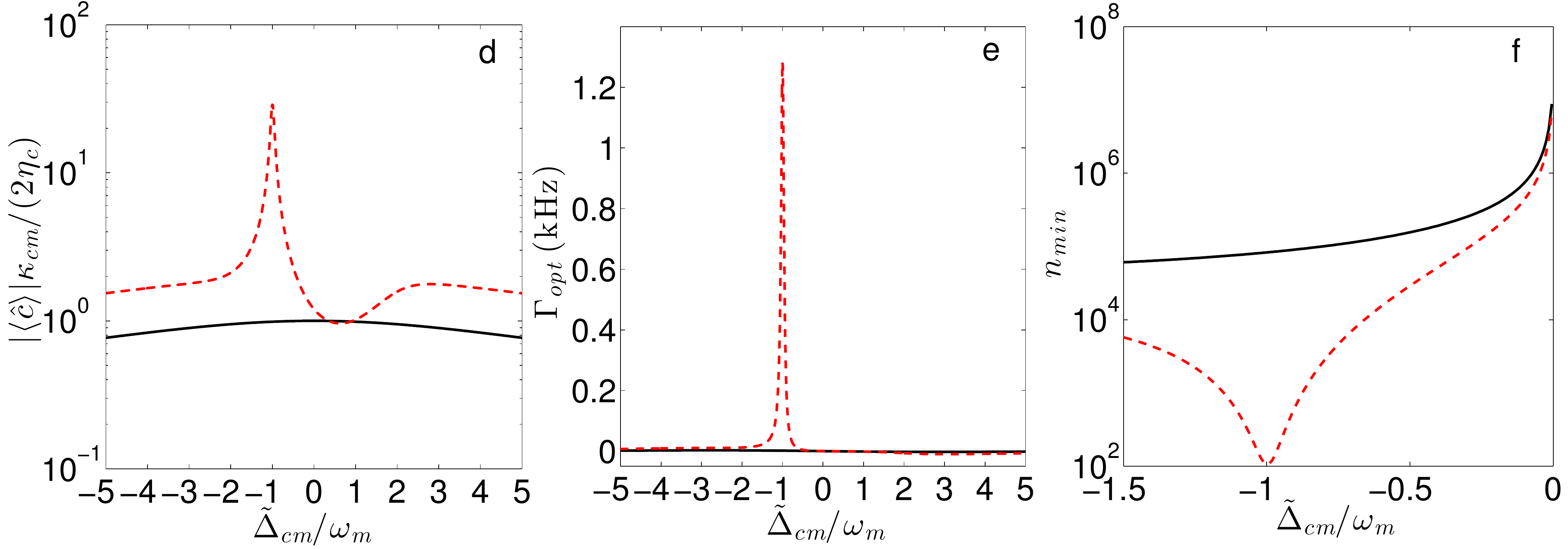}
\caption{(Color online). Features of the hybrid optomechanical cooling via RIR. Here $\omega_m = 2\pi\times 300$~kHz, $Q = 5 \times 10^7$, $T_{\rm bath} = 300$~K. $P = 1$~nW, and $g_0=2 \pi \times 200$~Hz. A $0.5$~mm atomic ensemble of $10^8$ atoms is illuminated with a control field of strength $\Omega = 2.6\gamma_e$ and $\Delta_a = -15 \gamma_e$. We show the intensity in the optomechanical cavity, the optical damping $\Gamma_{opt}$ and the minimum number of phonons for $\kappa_{cm} = 2\pi \times 240$~kHz (a,b,c) and $\kappa_{cm} = 2\pi \times 3.6$~MHz (d,e,f), respectively. Black solid lines: no cavity coupling, $J = 0$. Red dashed lines:  hybrid case.}
\end{center}
\label{fig:OMeffects}
\end{figure*}

One can envison a similar enhancement of optomechanical cooling using the motional states of the atomic gas via a recoil-induced resonance. In order to take advantage of the spectrally narrow gain feature associated with the RIR, we turn to a configuration where the amplified output from the atomic medium directly drives the optomechanical cavity. Furthermore, in order to ensure gain within a single frequency window, we assume that the atomic medium is trapped in free space instead of within a cavity (see Fig.~\ref{fig:coupling} B). 

To account explicitly for the effects of photon recoil is is useful to introduce the new bosonic annihiltion operations  $\hat{a}\rightarrow\hat{a}_p=(1/\sqrt{2})\hat{a}e^{ikz}$, where the factor of $1/\sqrt{2}$ accounts for quantization of the field in terms of running waves modes \ and $e^{ikz}$ is the phase of the propagating field along the $z$ axis. The coupling field Hamiltonian Eq.\ref{eq:HoptA} becomes then
\begin{equation}
H'_{\rm optA}=\hbar \omega_2 \hat{a}_p^\dagger \hat{a}_p+i\hbar (\eta_a\hat{a}_p^\dagger-\eta_a^* \hat{a}_p),
\end{equation}
and the coupling between the two field modes
\begin{equation}
H'_{AM}=\hbar J(\hat{a}_p^\dagger \hat{c}+\hat{a}_p\hat{c}^\dagger),
\end{equation}
where $J=\sqrt{\kappa_{a} \kappa_\mathrm{cm}/2}$ and $\kappa_a = c/L_a$ is the free-space decay rate of an atomic cloud of length $L_a$. A large decay rate implies that the expectation value of the coupling field comes to a steady state over a very short time period. Thus,
\begin{equation}
\langle \hat{a}_p\rangle \rightarrow \langle \hat{a}_p\rangle =\frac{\eta_a}{-i\chi+\kappa_a/2}.
\end{equation}
Inserting this form in the equation of motion for $\langle \hat{c} \rangle$, and in a frame rotating at $\omega_2$ gives
\begin{equation}
\langle \dot{\hat{c}}\rangle =\left(i\tilde{\Delta}_\mathrm{cm}-\frac{\kappa_\mathrm{cm}}{2}\right)\langle \hat{c}\rangle +(\eta_c-iJ\langle \hat{a}_p\rangle).
\end{equation} 

We now have an additional drive term that has a frequency dependence due to the susceptibility of the atomic cloud. As a result, the steady state and fluctuations of the optomechanical cavity field are given by
\begin{align}
&\langle \hat c \rangle=(\eta_c-iJ\langle \hat{a}_p\rangle)\left(-i\tilde{\Delta}_\mathrm{cm}+\frac{\kappa_\mathrm{cm}}{2}\right)^{-1},\\
&\langle \delta c[\omega]\rangle =-ig_0\langle \hat c \rangle\langle \delta x[\omega]\rangle \left(-i(\tilde{\Delta}_\mathrm{cm}+\omega)+\frac{\kappa_\mathrm{cm}}{2}\right)^{-1}.
\end{align}
The dynamical radiation pressure force at the mechanical frequency thus becomes
\begin{align}\nonumber
&\langle \delta \hat F_{RP}[\omega_m]\rangle =i\hbar (g^2/x_{\rm zpt}) \langle \delta \hat x[\omega_m] \rangle \times\\
&\left(\frac{1}{-i(\tilde{\Delta}_\mathrm{cm}+\omega_m)+\kappa_\mathrm{cm}/2}-\frac{1}{i(\tilde{\Delta}_\mathrm{cm}-\omega_m)+\kappa_\mathrm{cm}/2}\right),
\end{align}
where $g=g_0|\langle \hat c \rangle|^2$, and the intra-cavity field evaluated at $\omega=\omega_2+ \omega_m$. The optically mediated cooling rate and spring constant become
\begin{eqnarray}\nonumber
\Gamma_{\rm opt}&=& 2 g_0^2|\langle \hat c \rangle|^2\, {\rm Re}\left[\frac{1}{-i\tilde{\Delta}_\mathrm{cm}^++\kappa_\mathrm{cm}/2}
-\frac{1}{i\tilde{\Delta}_\mathrm{cm}^-+\kappa_\mathrm{cm}/2}\right],\\ \nonumber
k_{\rm opt}&=& 2 m \omega_m\,g_0^2 |\langle \hat c \rangle|^2\nonumber \\
&&{\rm Im}\left[\frac{1}{-i\tilde{\Delta}_\mathrm{cm}^++\kappa_\mathrm{cm}/2}
-\frac{1}{i\tilde{\Delta}_\mathrm{cm}^-+\kappa_\mathrm{cm}/2}\right],
\end{eqnarray}
where $\tilde{\Delta}_\mathrm{cm}^{\pm}=\tilde{\Delta}_\mathrm{cm}\pm\omega$. 

Figure 11 
shows the intracavity intensity, optomechanical cooling rates, and minimum steady state occupation number of the mechanical mode with and without the atomic medium. In these results, the parameters for the atomic ensemble and the coupling field are chosen so as to realize a gain feature around the mechanical resonance frequency $\omega_m$.  The optomechanical cavity parameters and the mechanical oscillator parameters are the same as for the EIT case, except for a lower input power of 1 nW to avoid parametric instabilities. As can be seen, the coupling to the atomic medium results in a dramatic enhancement to the cooling rate for $\tilde{\Delta}_{\rm cm} \approx - \omega_m$. It corresponds to a decrease in the phonon number of the mechanical resonator by over two orders of magnitude. 

\begin{figure}[h]
\begin{center}
\includegraphics[width =\columnwidth]{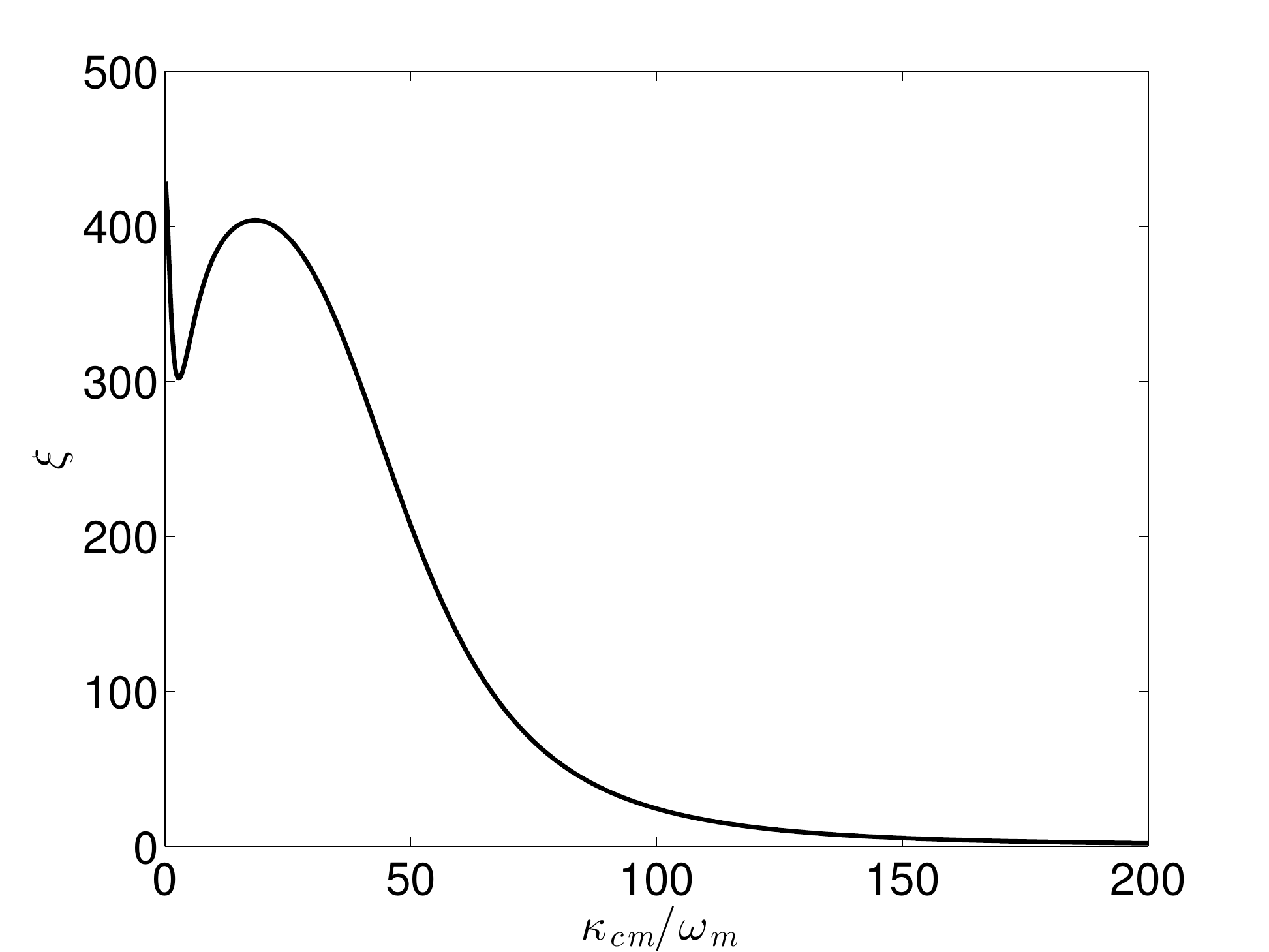}
\caption{Improvement factor for RIR-based hybrid optomechanical cooling with varying cavity damping rate. Here, $\omega_m = 2 \pi \times 300$~kHz, $Q = 5 \times 10^7$, $T_{bath} = 300$~K and $P = 1$~nW. The atomic parameters for the RIR are the same as in Fig. 11. }
\label{fig:RIRimprov}
\end{center}
\end{figure}

The enhancement to optomechanical cooling due to the RIR can be quantified in terms of the dimensionless parameter $\xi$ of Eq.~(\ref{eq:improvf}), see Fig.~\ref{fig:RIRimprov}. It reveals substantial optomechanical cooling due to the atomic medium over a wide region extending well into the Doppler regime of the optomechanical cavity. (Note that for very large $\kappa_{\rm cm}$, the optomechanical cavity becomes too lossy for an appreciable intensity to build up within the cavity, leading to a decreased influence of the atomic medium.)

\subsection{Comparison of the two coupling schemes}

While distinct physical mechanisms are at the origin of these results both the EIT-feedback coupling and the RIR-cascade coupling contribute to a substantial improvement of the mechanical cooling in the Doppler regime. 
In the case of EIT, the creation of a narrow window of field suppression around cavity resonance allows to eliminate the unwanted sideband thus improving cooling. This explains why cooling occurs on the blue side of the cavity resonance, in contrast to the more familiar resolved sideband regime situation. In the case of cascade coupling, the intrinsic asymmetry in the two sidebands is strongly enhanced as a result of interaction of the single-mode coupling field with the atomic medium, resulting in cooling on the red side of the cavity resonance.

A comparison of the improvement factors for the EIT and the RIR scheme, see Figs. 10 and 12, indicates that the latter yields the most dramatic improvement in optomechanical cooling. However, this comes at the expense of diminished tunability due to the sensitive dependence of the RIR process on trap parameters and the temperature of the atomic ensemble. 
 
\section{Conclusion}
In summary we have investigated two hybrid quantum systems consisting of a cavity optomechanical system optically coupled to an ultracold atomic gas. We demonstrated theoretically how the dispersive and gain optical properties of the atomic gases are exploited to modify the optomechanical response of the mechanical resonator, resulting in significantly enhanced cooling of the resonator, even for an optomechanical system that is nominally in the unresolved sideband regime.  We considered both the interaction of the optomechanical system with the spin degree of freedom of the atomic gas through a EIT feature as well as an interaction with its motional degree of freedom through a recoil induced resonance (RIR). In either case we found broad and robust parameter regimes wherein the mechanical resonator can be cooled to the ground state from room temperature. In the case of EIT, the improvement in optomechanical cooling is due to the narrow transparency window at the two-photon resonance that enhances the spectral asymmetry between the Stokes and anti-Stokes sidebands induced by mechanical motion. In the case of the RIR, optical gain enhances the anti-Stokes sideband leading to enhanced cooling. These results pave the way towards ground state optical cooling of low frequency mechanical resonators. 

The concrete examples considered here illustrate in relatively simple situations realizable with existing technology the considerable advantages provided by the exquisite optical control of ultracold atomic gases for the quantum control and manipulation of a mesoscopic mechanical resonators. They also hint at powerful schemes that can be conceived to dynamically tune the optical response of cavity optomechanical systems for various sensing, transduction and state transfer protocols. These aspects of hybrid systems with ultracold atoms will be considered in some detail in future work. 

\section{Acknowledgements}
We thank S. Steinke, Y. S. Patil, S. Chakram and S. Yelin for useful discussions. This work was supported by the DARPA QuASAR and ORCHID programs through grants from AFOSR and ARO, the U.S. Army Research Office,  the US NSF, the Cornell Center for Materials Research with funding from the NSF MRSEC program (DMR-1120296) and the NSF INSPIRE program. M. V. acknowledges support from the Alfred P. Sloan Foundation. F. B. dedicates this work to his wife Elizabeth and his son Paolo. 
\appendix

\section{Equations of motion for the RIR atomic operators}
Starting for the identity
\begin{equation}
\frac{d}{dt}\langle \hat{c}_j'^\dagger(k')\hat{c}_j(k)\rangle=\frac{d}{dt}\rho_{jj'}(k,k')=\frac{i}{\hbar}\langle [H, \hat{c}_j'^\dagger(k')\hat{c}_j(k)] \rangle.
\end{equation}
and the Hamiltonian (\ref{eq:aHamiltonian}), the Heisenberg equations for the atomic  operators are
\begin{widetext}
\begin{eqnarray}\nonumber
\frac{d}{dt}\rho_{gg}(k,k') &=&-\frac{i\hbar}{2 m_a}(k^2-k'^2)\rho_{gg}(k,k') +\Omega e^{i\omega_1 t}\rho_{eg}(k-k_1,k') \\
&&+\Omega e^{-i\omega_1 t}\rho_{ge}(k,k'-k_1) +\mathcal{E}_a \langle\hat{a}^\dagger\rangle e^{i\omega_2 t}\rho_{eg}(k-k_2,k') +\mathcal{E}_a \langle\hat{a}\rangle e^{-i\omega_2 t}\rho_{ge}(k,k'-k_2),\\ \nonumber
\frac{d}{dt}\rho_{ee}(k,k') &=&\left( \frac{i\hbar}{2 m_a}(k'^2-k^2)+i\omega_0\right)\rho_{ee}(k,k') +\Omega e^{i\omega_1 t}\rho_{eg}(k,k'-k_1) \\
&&+\Omega e^{-i\omega_1 t}\rho_{ge}(k-k_1,k) +\mathcal{E}_a \langle\hat{a}^\dagger\rangle e^{i\omega_2 t}\rho_{eg}(k,k'-k_2) +\mathcal{E}_a \langle\hat{a}\rangle e^{-i\omega_2 t}\rho_{ge}(k-k_2,k'),\\\nonumber
\frac{d}{dt}\rho_{eg}(k,k') &=&-i\omega_0\rho_{eg}(k,k') +\Omega e^{-i\omega_1 t}\left( \rho_{ee}(k,k'+k_1)-\rho_{gg}(k-k_1,k')\right) \\
&&+\mathcal{E}_a \langle\hat{a}\rangle e^{-i\omega_2 t}\left(\rho_{ee}(k,k'+k_2) +\rho_{ge}(k-k_2,k')\right).
\end{eqnarray}
\end{widetext}
In a frame rotating at the control frequency,$\omega_1$, so that $\tilde{\rho}_{eg}(k,k')=\rho_{eg}(k,k')e^{i\omega_1 t}$, and with  $\Delta_a = \omega_0-\omega_1$, and $\delta=\omega_2-\omega_1$, we have then
\begin{eqnarray}\nonumber
&&\frac{d}{dt}\tilde{\rho}_{eg}(k,k')=-i\Delta_a \tilde{\rho}_{eg}(k,k') \\ \nonumber
&&+\Omega \left( \rho_{ee}(k,k'+k_1)-\rho_{gg}(k-k_1,k')\right) \\
&&+\mathcal{E}_a \langle\hat{a}\rangle e^{-i\delta t}\left(\rho_{ee}(k,k'+k_2) +\rho_{ge}(k-k_2,k')\right).
\end{eqnarray}
The steady state solution for this coherence is  
\begin{eqnarray}\nonumber
&&\tilde{\rho}_{eg}(k,k') =-i\frac{\Omega}{\Delta_a}\left( \rho_{ee}(k,k'+k_1)-\rho_{gg}(k-k_1,k')\right) \\
&-&i\frac{\mathcal{E}_a \langle\hat{a}\rangle}{\Delta_a} e^{-i\delta t}\left(\rho_{ee}(k,k'+k_2) +\rho_{gg}(k-k_2,k')\right).
\end{eqnarray}
Here we have used the assumption that $\Delta_a \pm \hbar (k^2-k'^2)/(2m)\approx \Delta_a$, since the control-atomic resonance detuning ($\Delta_a$) is much larger than the atomic recoil frequency.
We shall use this expression in the equation of motion for the fields, and coherences between the same electronic state of the atomic ensemble. We now eliminate the excited state population and coherences, $\rho_{ee}\rightarrow 0$, since the spontaneous emission rate is much larger than the rate of population and coherence buildup, $\gamma_{e} \gg |\Omega|^2/\Delta_a$.

Introducing the new variables introduced in main tex, such as e.g. the recoil wave vector $2 k_0=k_1-k_2$, dimensionless momentum $p=\hbar k/2 \hbar k_0$ and a dimensionless parameter characterizing control coupling $\beta = \Omega/\Delta_a$, we can rewrite the equation of motion for the ground state using the steady state values for coherences between the ground states as 
\begin{eqnarray}\nonumber
&&\dot{\rho}_{gg}(p,p')=4i\omega_r(p'^2-p^2)\rho_{gg}(p,p')\\ \nonumber
&& +i\beta \mathcal{E}_a \langle\hat{a}\rangle e^{-i \delta t} \left[ \rho_{gg}(p+1,p') -\rho_{gg}(p,p'-1)\right]\\
&& -i\beta \mathcal{E}_a \langle\hat{a}^\dagger\rangle e^{i \delta t}\left[ \rho_{gg}(p,p'+1) -\rho_{gg}(p-1,p')\right].
\label{eq:gsceom}
\end{eqnarray}
Equations  (\ref{eq:gsceom}) are then used to derive explicit forms for the population and coherences in the ground state, see Eq. (\ref{eq:gseom1}). Assuming that the atomic populations remain in a thermal state and that the coherences come to a steady state over the time scale of the evolution of the electric fields,  we finlly obtain
 \begin{eqnarray}\nonumber
\zeta_{p-1,s}&=&\frac{i\beta^*\mathcal{E}_a^* N(\Pi_{p-1}-\Pi_p)}{\left(4i\omega_r(2p-1)+i\delta-\gamma_{\rm coh}\right)} \langle\hat{a}\rangle,\\ \nonumber
\zeta_{p+1,s}&=&-\frac{i\beta \mathcal{E}_a N(\Pi_{p+1}-\Pi_p)}{\left(4i\omega_r(2p+1)+i\delta+\gamma_{\rm coh}\right)}\langle\hat{a}^\dagger\rangle.
 \end{eqnarray}
These are the steady state values of the atomic coherences used in the equation of motion of the coupling field.



\begin{thebibliography}{10}

\bibitem{OM_review} M. Aspelmeyer, T. Kippenberg and F. Marquardt, arXiv:1303.0733 (2013); P. Meystre, Annalen der Physik {\bf 525}, 215 (2013).

\bibitem{OM_cooling} F. Marquardt, J. P. Chen, A. A. Clerk, and S. M. Girvin, Phys. Rev. Lett. {\bf 99}, 093902 (2007);I. Wilson-Rae, N. Nooshi, W. Zwerger, and T. J. Kippenberg, Phys. Rev. Lett. {\bf 99}, 093901 (2007).

\bibitem{Cleland}A. D. O'Connell, M. Hofheinz, M. Ansmann, R. C. Bialczak, M. Lenander, E. Lucero, M. Neeley, D. Sank, H. Wang, M. Weides, J. Wenner, J. M. Martinis, and A. N. Cleland, Nature  (London)  \textbf{464}, 697 (2010).

\bibitem{Teufel}J. D. Teufel,  T. Donner,  Dale Li,  J. W. Harlow,  M. S. Allman,  K. Cicak,  A. J. Sirois,  J. D. Whittaker,  K. W. Lehnert, and R. W. Simmonds, Nature  (London) \textbf{475}, 359 (2011).

\bibitem{Painter}J. Chan, T. P. Mayer Alegre, A. H. Safavi-Naeini, J. T. Hill, A. Krause, S. Gr$\ddot{\rm{o}}$blacher, M. Aspelmeyer, and O. Painter, Nature  (London) \textbf{478}, 89 (2011); A. H. Safavi-Naeini, J. Chan, J. T. Hill, T. P. Mayer Alegre, A. Krause, and O. Painter, Phys. Rev. Lett. \textbf{108}, 033602 (2012).

\bibitem{hybrid2014}
B. Rogers, N. Lo Gullo, G. De Chiara, G. M. Palma, and M. Paternostro, arXiv:1402.1195, (2014).

\bibitem{PKU}
Yong-Chun Liu, Rui-Shan Liu, Chun-Hua Dong, Yan Li, Qihuang Gong, and Yun-Feng Xiao, arXiv:1406.7359 (2014).

\bibitem{atom_cavity}
H. Ritsch, P. Domokos, F. Brennecke, and T. Esslinger, Rev. Mod. Phys. {\bf 85}, 553 (2013).

\bibitem{treutlein-genes} P. Treutlein, C. Genes, K. Hammerer, M. Poggio, and P. Rabl, arXiv:1210.4151 (2012).

\bibitem{bowen} J.S. Bennet, L.S. Madsen, M. Baker, H. Rubinsztein-Dunlop, and W.P. Bowen, arXiv:1404.3445 (2014).

\bibitem{genes} A. Dantan, B. Nair, G. Pupillo, and C. Genes, arXiv:1406.7100 (2014).

\bibitem{EIT_review}
M. Fleischhauer, A. Imamoglu, and J.P. Marangos, Rev. Mod. Phys. {\bf 77}, 633 (2005).

\bibitem{Grynberg1992}
J. Guo, P.R. Berman, B. Dubetsky, and G. Grynberg, Phys. Rev. A {\bf 46}, 1426 (1992).

\bibitem{Grynberg1994}
J.-Y. Courtois, G. Grynberg, B. Lounis, and P. Verkerk, Phys. Rev. Lett. {\bf 72}, 3017 (1994).

\bibitem{Moore1998}
M. G. Moore and P. Meystre, Phys. Rev. A {\bf 58}, 3248 (1998).

\bibitem{Vengalattore2005}
M. Vengalattore and M. Prentiss, Phys. Rev. A {\bf 72}, 021401(R) (2005).

\bibitem{Hafezi2008}
M. Vengalattore, M. Hafezi, M. Lukin and M. Prentiss, Phys. Rev. Lett. {\bf 101}, 063901 (2008).

 \bibitem{Chakram2014}
 S. Chakram, Y. S. Patil, L. Chang and M. Vengalattore, Phys. Rev. Lett. {\bf 112}, 127201 (2014).
 
 \bibitem{Genes2009}
C. Genes, H. Ritsch, and D. Vitali, Phys. Rev. A {\bf 80}, 061803(R) (2009).

\bibitem{Genes2011}
C. Genes, H. Ritsch, M. Drewsen and A. Dantan, Phys. Rev. A {\bf 84}, 051801(R) (2011).
 
\bibitem{kappa}
For a single-sided cavity, meaning one with a completely reflecting (right) end-mirror, we have $\kappa_\mathrm{l,ca} = \kappa_\mathrm{ca}$, while for a cavity with equal mirrors at both ends, we have $\kappa_\mathrm{l,ca} = \kappa_\mathrm{r,ca} = \kappa_\mathrm{ca}/2$.

\bibitem{GardinerZollerbook}
 C.W. Gardiner and P. Zoller, {\it Quantum Noise} (Springer, Berlin, 2004).
  
 \bibitem{Patil2014}
 Y. S. Patil, S. Chakram. L. Aycock and M. Vengalattore, arXiv:1404.5583 (2014).

\end{thebibliography}
\end{document}